\documentclass[usenatbib,a4paper,usenames,dvipsnames]{mn2e}
\usepackage{amsmath}
\usepackage{amssymb}
\usepackage{color}
\definecolor{darkblue}{rgb}{0.0,0.0,0.3}
\usepackage{hyperref}
\hypersetup{colorlinks,breaklinks,
           linkcolor=darkblue,urlcolor=darkblue,
           anchorcolor=darkblue,citecolor=darkblue}

\usepackage{txfonts}

\DeclareSymbolFont{cmletters}{OML}{cmm}{m}{it}
\DeclareMathSymbol{v}{\mathalpha}{cmletters}{"76}
\usepackage{graphicx}

\newcommand{\useiop}{-3}

\newcommand{\shortauthors}[1]{}
\newcommand{\shorttitle}[1]{}
\newcommand{\altaffiltext}[2]{}
\newcommand\aj{\rmfamily{AJ}}%
\newcommand\apj{\rmfamily{ApJ}}%
\newcommand\apjl{\rmfamily{ApJ}}%
\newcommand\apss{\rmfamily{Ap\&SS}}%
\newcommand\aap{\rmfamily{A\&A}}%
\newcommand\mnras{\rmfamily{MNRAS}}%
\newcommand\pasj{\rmfamily{PASJ}}%
\newcommand\nat{\rmfamily{Nature}}%
\usepackage{ifthen}

\defcitealias{tch08}{TMN08}
\defcitealias{tch09}{TMN09}
\defcitealias{tch11}{TNM11}
\defcitealias{kom09}{K09}
\defcitealias{bz77}{BZ}
\defcitealias{bp82}{BP}

\newcommand{\OmegaH}{\Omega_\mathrm{H}}
\newcommand{\PhiH}{\Phi_\mathrm{BH}}

\newcommand{\Mdot}{\dot M}

\newcommand{\PhiBH}{\Phi_{\rm BH}}
\newcommand{\MBH}{M_{\rm BH}}

\newcommand{\cut}[1]{\hbox{}}

\newcommand{\Gcmsq}{\ensuremath{{\rm G\,cm^2}}}

\shortauthors{}
\shorttitle{}

\ifthenelse{\equal{\useiop}{-3}}{ 

\author[A.~Tchekhovskoy 
and
D.~Giannios]
{Alexander Tchekhovskoy$^{1,2}$\thanks{\hbox{E-mail:
      atchekho@berkeley.edu~(AT), dgiannio@purdue.edu (DG)}} and
Dimitrios Giannios$^3$
\\
 $^1$Department of Physics and Department of Astronomy, University of California, Berkeley, CA 94720-3411\thanks{Einstein Fellow}\\
 $^2$Lawrence Berkeley National Laboratory, 1 Cyclotron Rd, Berkeley,
  CA 94720, USA\\
 $^3$Department of Physics and Astronomy, Purdue University, 525 Northwestern Avenue, West Lafayette, IN 47907, USA}
}{
}

\begin{document}
\label{firstpage}

\title[Magnetic Flux of Progenitor Stars Sets Gamma-ray Burst
Luminosity and Variability]%
{Magnetic Flux of Progenitor Stars Sets Gamma-ray Burst
Luminosity and Variability}

\ifthenelse{\equal{\useiop}{-2}}{
\journal{New Astronomy}
\begin{frontmatter}

\author[cfa]{Alexander Tchekhovskoy}
\ead{atchekho@princeton.edu}

\address[cfa]{Princeton Center for Theoretical Science, Jadwin Hall, Princeton
  University, Princeton, NJ 08544, USA}
}{}
\ifthenelse{\equal{\useiop}{3}}{
\author{Alexander Tchekhovskoy$^1$ and $^2$} 
  \maketitle
  \begin{affiliations}
    \item Princeton Center for Theoretical Science, Princeton
  University, Jadwin Hall, Princeton
  NJ 08544; atchekho@princeton.edu\\
  \end{affiliations}
}{}
\ifthenelse{\equal{\useiop}{0}}{self-consistent 
\altaffiltext{1}{Princeton Center for Theoretical Science, Princeton
  University, Jadwin Hall, Princeton
  NJ 08544; atchekho@princeton.edu}
}

\ifthenelse{\equal{\useiop}{-3}}{ 
\date{Accepted . Received ; in original form }
\pagerange{\pageref{firstpage}--\pageref{lastpage}} \pubyear{2012}
\maketitle
}

\begin{abstract}
  Long-duration gamma-ray bursts (GRBs) are thought to come from the
  core-collapse of Wolf-Rayet stars. Whereas their stellar
  masses~$M_*$ have a rather narrow distribution, the population of
  GRBs is very diverse, with gamma-ray luminosities~$L_\gamma$ spanning
  several orders of magnitude.  This suggests the existence of a
  ``hidden'' stellar variable whose burst-to-burst variation leads to
  a spread in~$L_\gamma$.  Whatever this hidden variable is, its variation
  should not noticeably affect the shape of GRB lightcurves, which
  display a constant luminosity (in a time-average sense) followed by
  a sharp drop at the end of the burst seen with
  \emph{Swift}/XRT.  We argue that such a hidden variable is
  progenitor star's large-scale magnetic flux. Shortly after the core
  collapse, most of stellar magnetic flux accumulates near the black
  hole (BH) and remains there. The flux extracts BH rotational energy
  and powers jets of roughly a constant luminosity,~$L_j$. However,
  once BH mass accretion rate $\dot M$ falls below $\sim L_j/c^2$, the
  flux becomes dynamically important and diffuses outwards, with the
  jet luminosity set by the rapidly declining mass accretion rate,
  $L_j\sim \Mdot c^2$.  This provides a potential explanation for the
  sharp end of GRBs and the universal shape of their
  lightcurves. During the GRB, gas infall translates spatial variation
  of stellar magnetic flux into temporal variation of $L_j$. We make
  use of the deviations from constancy in~$L_j$ to perform stellar
  magnetic flux ``tomography''. Using this method, we infer the
  presence of magnetised tori in the outer layers of progenitor
  stars for GRB~920513 and GRB~940210.
\end{abstract}

\ifthenelse{\equal{\useiop}{-2}}{
\end{frontmatter}
}{}

\ifthenelse{\equal{\useiop}{-3}}{ 
}{}
\ifthenelse{\equal{\useiop}{3}}{
  }{}
  
\ifthenelse{\equal{\useiop}{0}}{
{
  }
}


\section{Introduction}
\label{sec:introduction}

Long-duration GRBs are believed to be associated with 
the collapse of the core of massive stars. This association is 
very firm in GRBs accompanied by supernovae explosions
\citep{stanek03,galama98}. 
The nature of the supernova (type Ic) indicates 
Wolf Rayet stars of spectral type WO/C as the progenitors. Such stars
have mass $\sim 10M_\odot$ and radius of $R\sim$ a few
$R_\odot$. Despite their masses have a rather narrow distribution,
the GRBs they produce come in a very broad range of power. Their luminosity
function extends over at least 4 orders of magnitude in $L_\gamma$
\citep{2010MNRAS.406.1944W}, and various additional categories of GRBs
have emerged in recent years \citep[e.g.,][]{2014ApJ...781...13L}.  

The observed duration of the 
prompt emission episode ($\sim1-100$ sec) roughly 
agrees with the free-fall time scale of the progenitor
star. The GRB emission is notoriously variable;
composed of a large number of intense gamma-ray pulses.
Despite their erratic behaviour, GRBs have some well-defined 
properties which tightly constrain the models for the central
engine. The time-averaged properties of the prompt emission 
do not evolve in a systematic way with time since the trigger 
\citep{2000ApJ...539..712R}. 
By looking at a random segment of the GRB lightcurve there is no way to tell
from the amplitude and duration of the pulses and the intervals
between the pulses 
whether the segment corresponds to the first or second half of the GRB
(\citealt{2000ApJ...539..712R,2002A&A...385..377Q}).
This is also demonstrated by the constant slope $S$ of cumulative counts 
during GRBs revealing that $\int_0^t L_\gamma{\rm d}t\sim {\rm
  constant}\times t$ or $L_\gamma\sim$ constant  
\citep{2002A&A...393L..29M}.
In contrast, the end of the GRB is typically well-defined and is  marked 
by a steep decline in flux. This end stage is seen
in X-rays with {\it XRT} on {\it Swift}, and is characterised by a
steep time-dependence of luminosity, with the flux dropping 
by several orders of magnitude. This abrupt drop is consistent with 
an abrupt turn-off of the central engine (see
\citealt{2006ApJ...642..389N,2006JPhCS..46..403Z}). 

This GRB behaviour is hard to understand in the context of core-collapse
models irrespective of whether GRBs are powered by 
accretion onto a black hole (BH, \citealt{woosley_gamma_ray_bursts_1993}) or by the rotation of a
neutron star \citep{usov_magnetar_grbs_1992}. 
How can the average luminosity of the jet,  for
a minute or so, ignore central engine evolution 
when the latter is expected to evolve substantially on a similar
timescale? What marks the sudden decline in the GRB emission? What
sets the GRB luminosity and what causes it to differ so much from burst to burst?

\begin{figure}
\begin{center}
    \includegraphics[width=0.9\columnwidth]{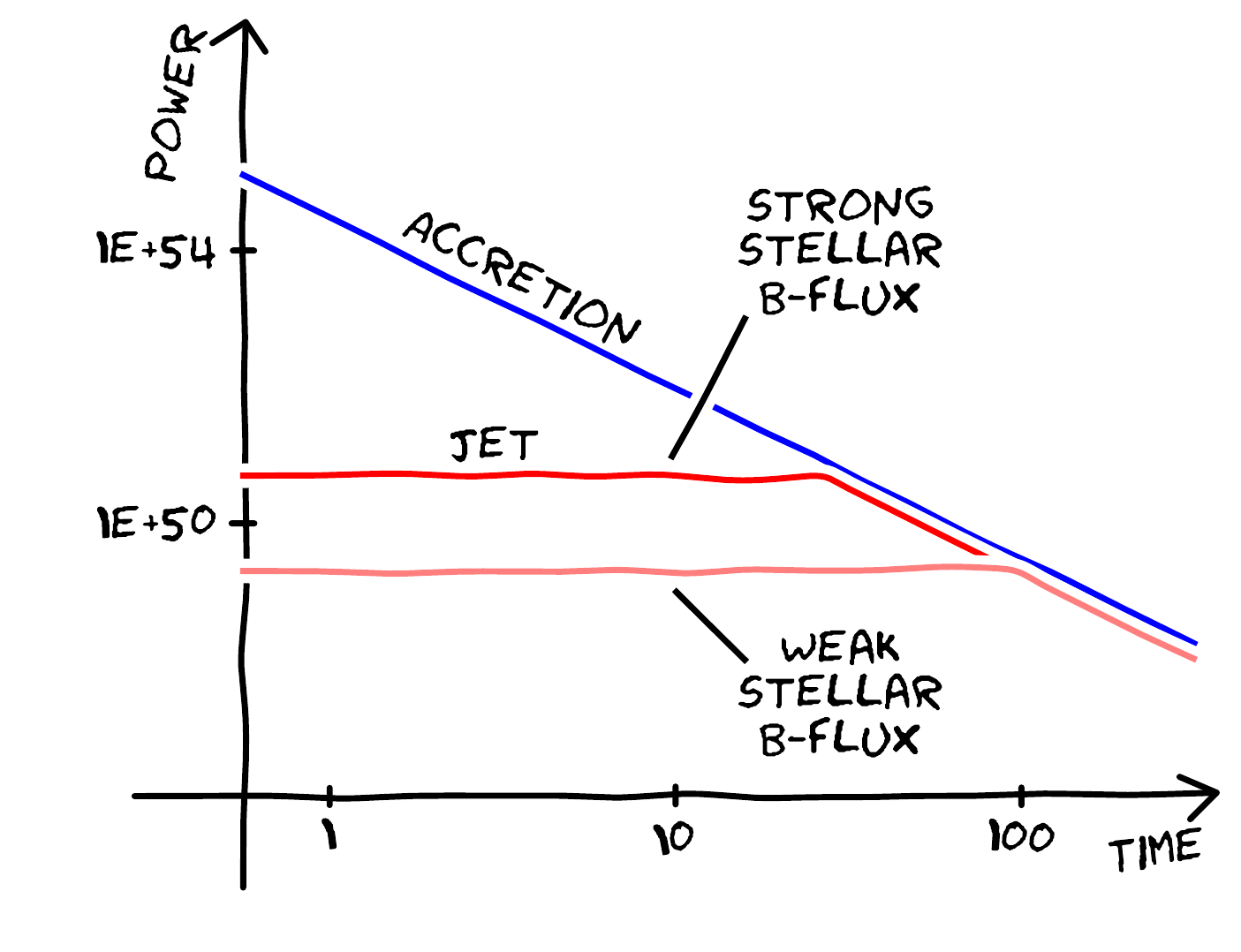}
  \end{center}
  \caption{A cartoon depiction of accretion and jet power
    time-dependence during a core collapse. The indicated values along
    the axes are approximate and intended to give a rough idea of
    characteristic energy and time scales.  At an early time,
    $t\lesssim$ few seconds, mass accretion rate is extremely high,
    $\Mdot c^2 \sim M_\odot c^2$~s$^{-1} \sim
    10^{54}$~erg~s$^{-1}$. Soon after core collapse, infalling gas
    drags into the BH most of progenitor star's large-scale magnetic
    flux, $\Phi_*$.  This flux sets BH jet power, $L_j \sim
    10^{50}$~erg~s$^{-1}\propto \Phi_*^2$, which remains approximately
    constant and \emph{independent of} $\Mdot$. That $L_j \lll \Mdot
    c^2$ means that BH magnetic flux is dynamically subdominant and
    easily held on the BH by the pressure of the accretion
    flow. However, $L_j$ cannot maintain its constancy indefinitely:
    if jet power remained constant, it would eventually exceed
    accretion power!  But this is impossible: if $L_j$ substantially exceeds $\Mdot
    c^2$, BH magnetic flux becomes dynamically-important and cannot be
    held on the BH by accretion \citep{tch11}. As a result, BH
    magnetic flux diffuses outward, such that jet luminosity tracks
    the rapidly decaying accretion power, $L_j\simeq \Mdot c^2$.  This
    steep decay of jet luminosity marks the end of the GRB. The
    stronger the stellar magnetic flux, the more luminous is the GRB:
    dark and light red lines illustrate the cases of stronger and
    weaker stellar magnetic fluxes, respectively.}
\label{fig:sketch}
\end{figure}

In the magnetar model of GRBs the jet luminosity can be 
constant  for $\sim 1$~min  
(set by the spin-down timescale of the magnetar). However, it is unclear
what causes GRBs to abruptly end.  Moreover,
jet properties (baryon loading) evolve fast with time
while the average observed GRB spectra do not (see 
\citealt{2011MNRAS.413.2031M} for a related discussion).

If the central compact object is a BH, 
accretion can power GRBs either 
through neutrino annihilation \citep{1999ApJ...518..356P,1999A&A...344..573R,2007A&A...463...51B,chen_neutrino_cooled_2007,2011MNRAS.410.2302Z} or magnetic energy 
extraction (\mbox{\citealt*{1992ApJ...395L..83N}};
\mbox{\citealt{le03}}; \mbox{\citealt{mr97}}). After the core collapse, the accretion rate $\Mdot$
drops fast with time unless one makes very particular choices for
the progenitor star and the accretion disk viscosity \citep{2008MNRAS.388.1729K,2008Sci...321..376K}. In any jet
production mechanism, which depends on the instantaneous BH mass accretion rate, 
$\Mdot$, the observed constancy of the GRB
luminosity requires fine tuning: in order to obtain an approximately
constant jet luminosity, the efficiency of jet production
must increase with time in exactly the right way so as to precisely cancel
the effect of decrease in $\Mdot$.
 
This monotonic decrease of $\dot M$ with time challenges neutrino models
where the jet power scales with accretion rate as $L_j\propto\dot
M^{9/4}$ (or a similar steep power; see
\citealt{2011MNRAS.410.2302Z}). Even a weak systematic change of the average $\dot M$
throughout the GRB should be easily detectable, which is not the case in
the majority of the bursts. The mechanism also cannot account for the
energetics of the longest bursts observed \citep{2014arXiv1408.4509L}. 

Here we argue that it is the \emph{magnetic flux through the collapsing star
that determines the luminosity of the jet}, as we illustrate in Fig.~\ref{fig:sketch}. Indeed, if the jet is launched via 
magnetic fields, the accretion rate does not always determine 
the jet power directly (\citealt*{tch11}; see however
\citealt{KrolikPiran2011}). The available magnetic flux in the
progenitor star may be much 
more constraining \citep{kb09}.
The flux accumulates on the newly
formed BH on a timescale shorter than the time it takes for
the jet to emerge from, or break through, the collapsing star. 

We assume that the jet 
is powered by the large-scale magnetic flux via the Blandford-Znajek process (BZ, \citealt{bz77}). In this picture, as in most collapsar models, the
jet breakout marks the start of the prompt emission, or the GRB
trigger. From then on and until the end of the GRB, for reasonable collapsar
parameters, the mass of the BH $\MBH$, the magnetic flux through
the hole $\Phi_{\rm BH}$, and BH spin $a$ evolve little over the
GRB duration. So long as the mass accretion rate $\dot M$ is sufficiently high to 
sustain the magnetic flux $\PhiBH$ on the BH, jet luminosity $L_j\propto
a^2 \Phi_{\rm BH}^2M_{\rm BH}^{-2}$
is \emph{independent of} $\dot M$ and is approximately constant, as
illustrated in Fig.~\ref{fig:sketch} with horizontal segments of red
lines. However, since $\Mdot$
asymptotically approaches zero, eventually $\Mdot$ becomes too low
to confine the magnetic flux of a constant strength on the BH.
This
occurs at the equipartition between accretion and jet powers,
$\dot Mc^2\simeq L_j\sim 10^{50}\,{\rm erg\ s^{-1}}$, or, in terms of
mass accretion rate, $\Mdot \sim 10^{-4}\ M_\odot$~s$^{-1}$. From this point on the
accretion disk cannot hold the entirety of magnetic field on the BH
anymore. The remaining flux on the BH stays in equipartition with
accretion power, and the jet
luminosity scales linearly with $\Mdot$, which rapidly declines toward zero and
causes the GRB to end abruptly.

We start with the description of progenitor star's structure and its core-collapse in
Sec.~\ref{sec:timing}, paying particular attention to how the central
BH properties --  mass, spin, and magnetic flux -- change in time.
In Sec.~\ref{sec:evol-jet-lumin} we derive the time-evolution of jet
power and in Sec.~\ref{sec:comparison-with-grb} perform the
comparison of our model to the observed GRB light curves. In
Sec.~\ref{sec:conclusions} we conclude.
Throughout the paper, we use Gaussian-cgs units.

\section{Collapsar models}
\label{sec:timing}

\begin{figure}
\begin{center}
    \includegraphics[width=0.9\columnwidth]{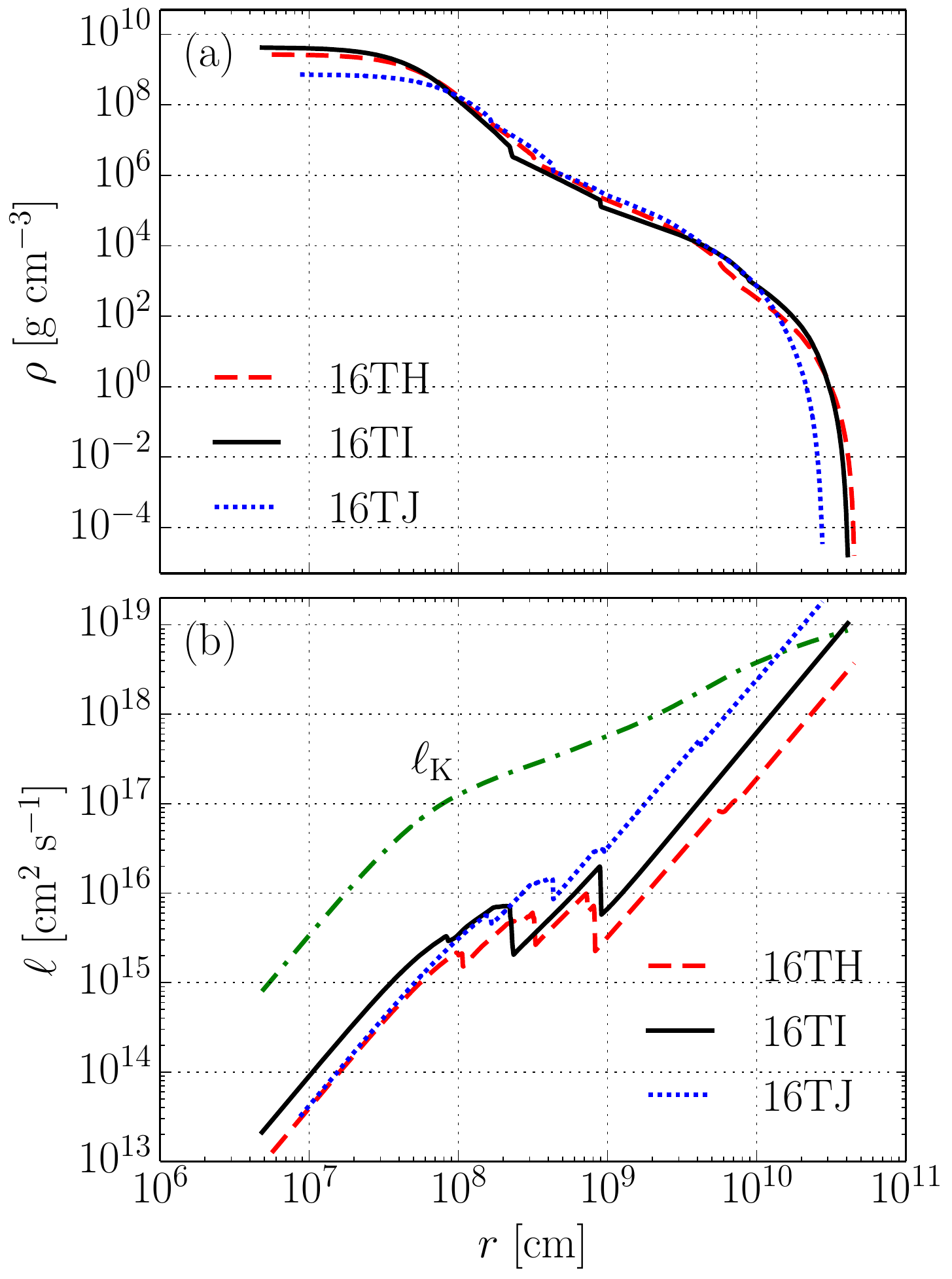}
  \end{center}
  \caption{Structure of GRB stellar progenitor models
    \citep{woosley_progenitor_2006}, 16TH, 16TI, and 16TJ, which are
    shown with red dashed, black solid, and blue dotted lines (see
    legend). {\bf [Panel a]:} Radial profile of stellar density,
    $\rho$.  Density is approximately constant inside the stellar core,
    $r\lesssim10^8$~cm, and falls off approximately as a power-law in radius
    outside of the core, $\rho\propto r^{-2.5}$, until the surface of
    the star, $r_*\sim {\rm few}\times 10^{10}$~cm, where it
    exponentially drops. {\bf [Panel b]:} Radial profile of specific
    angular momentum for the three models, which are shown with the
    same colours as in panel (a). For comparison, we also show the
    Keplerian angular momentum, $\ell_{\rm K}$, with green dash-dotted
    line. Discontinuities in the radial distribution reflect shells in
    stellar structure. Near the surface of the star, the angular
    momentum exceeds the Keplerian value.$^{\ref{fn:1}}$}
\label{fig:rhoell}
\end{figure}

We consider GRB jets that are powered by the accretion onto the central
BH.  To compute the rate at which the BHs are fed with
gas, we consider several pre-collapse stellar models described in
\citet{woosley_progenitor_2006}. We focus in this section on the
``16TI'' model of a pre-collapse star at  $1$\% solar metallicity. The
model includes a treatment of magnetic dynamo effects and mass and
angular momentum loss via stellar winds. 
Figure~\ref{fig:rhoell}(a) shows the density profile in the
pre-collapse star: outside the core, $R\gtrsim10^8$~cm, the density
falls off roughly as a power-law, $\rho \propto r^{-2.5}$, and cuts off
exponentially toward the surface, at $r_*\simeq5\times10^{10}$~cm.
Figure~\ref{fig:rhoell}(b) shows the radial distribution of angular
momentum.  It has clear discontinuities, which result from transitions
between different shells of stellar structure.  To get a sense of the
variety in GRB progenitors, we also consider two other
progenitor models, 16TH and 16TJ, which differ from the fiducial model
by the strength of stellar winds, and we
show them in Fig.~\ref{fig:rhoell} for comparison. As we will see
below, our main conclusions do not depend on the details of model (see
Section \ref{sec:effect-prec-model}).\footnote{\label{fn:1}Note that
  stellar rotation profile is uncertain due to the limitations of the
  1D stellar models and uncertainties in the treatment of microphysics
  of turbulence, magnetic fields, and stellar winds. For instance, the angular
  momentum $\ell = (2/3)\Omega r^2$, in models 16TH and 16TJ exceeds
  the Keplerian value, $\ell_{\rm K}$, in the outer layers of the
  star (see Fig.~\ref{fig:rhoell}b). If the outer layers indeed were
  super-Keplerian, the star would have already expelled these outer layers long before core collapse.  These extremely high values appear due to
  the uncertainties of angular momentum transport inside the outer layers
  of the star and angular momentum extraction 
  by the stellar winds and are not physical
  \citep{woosley_progenitor_2006}. For this reason, for our
  calculations, we modify the rotational profile of the models and
  limit the angular momentum to be at most $10$\% of the local
  Keplerian value (see Fig.~\ref{fig:timescales}(a) for an example; 
  our results are not sensitive to the particular
  value of the cutoff). Apart from this modification, we adopt the
  rotational profile of the models as is, to illustrate the robustness
  of our results.}

\begin{figure}
\begin{center}
    \includegraphics[width=1.05\columnwidth]{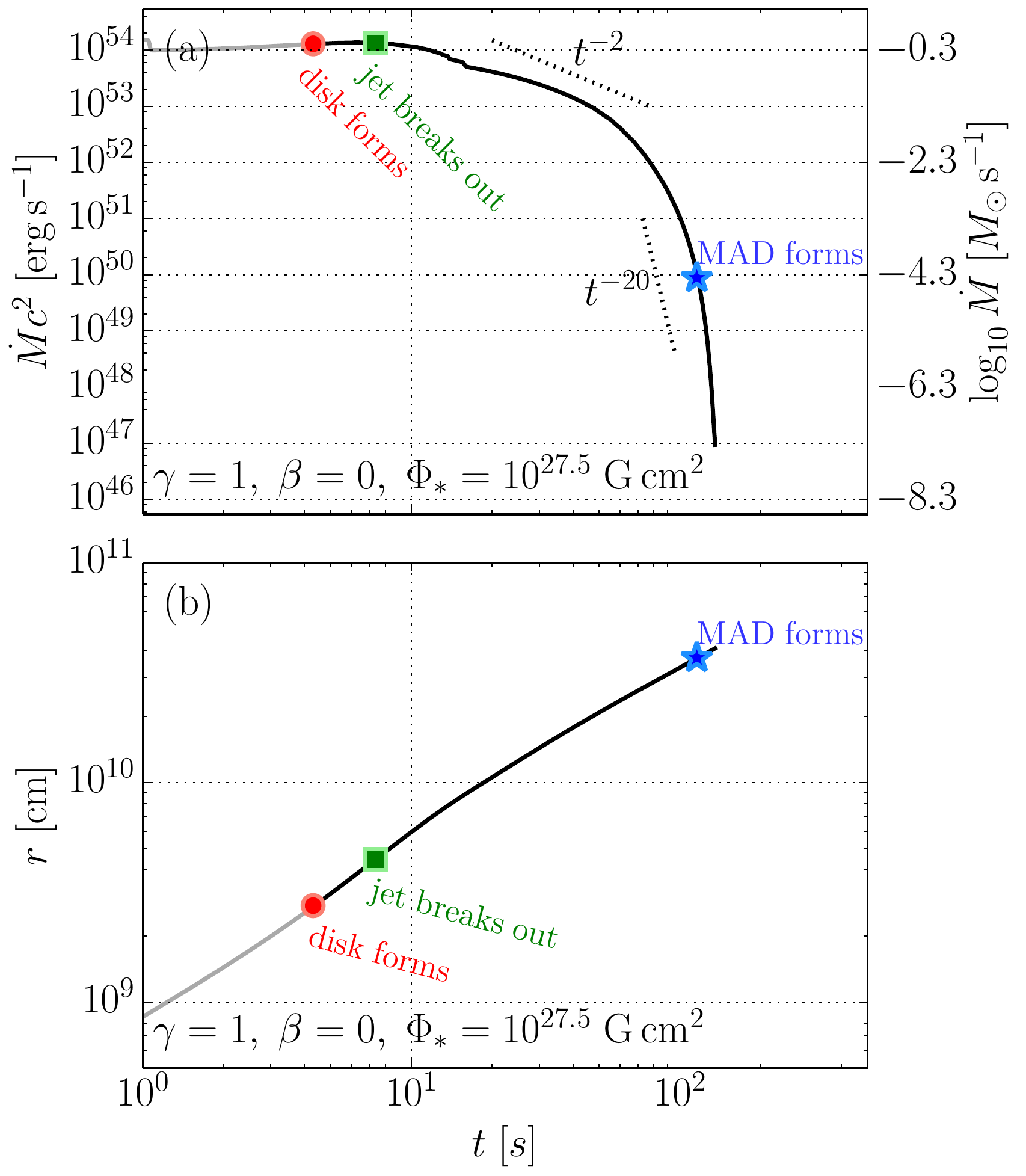}
  \end{center}
  \caption{{\bf [Panel (a)]:} Mass accretion rate $\dot M$ vs. time
    since core collapse $t$ for our fiducial pre-collapse stellar
    progenitor model, 16TI \citep{woosley_progenitor_2006}. For the
    sake of later discussion, the time of disk formation is
    indicated with a red circle, jet breakout with the green square,
    and emergence of a dynamically-important magnetic field
    threading the BH (or the magnetically-arrested disk, MAD), with
    the blue star, at which the GRB ends (see
    Sec.~\ref{sec:estimates}). The latter two of these times depend on progenitor star's
    magnetic flux strength (described by $\Phi_*$) and radial
    distribution (described by $\gamma$, see eq.~\ref{eq:PhivsM}), and
    the mass-loss from an accretion disk (described by
    $\beta$ see eq.~\ref{eq:beta}). {\bf [Panel (b)]:} The mapping between the
    the radial position $r$ of a layer of gas in the progenitor star
    and the time $t$ since core collapse at which it reaches the BH.}
\label{fig:mdot}
\end{figure}

\begin{figure}
\begin{center}
    \includegraphics[width=0.9\columnwidth]{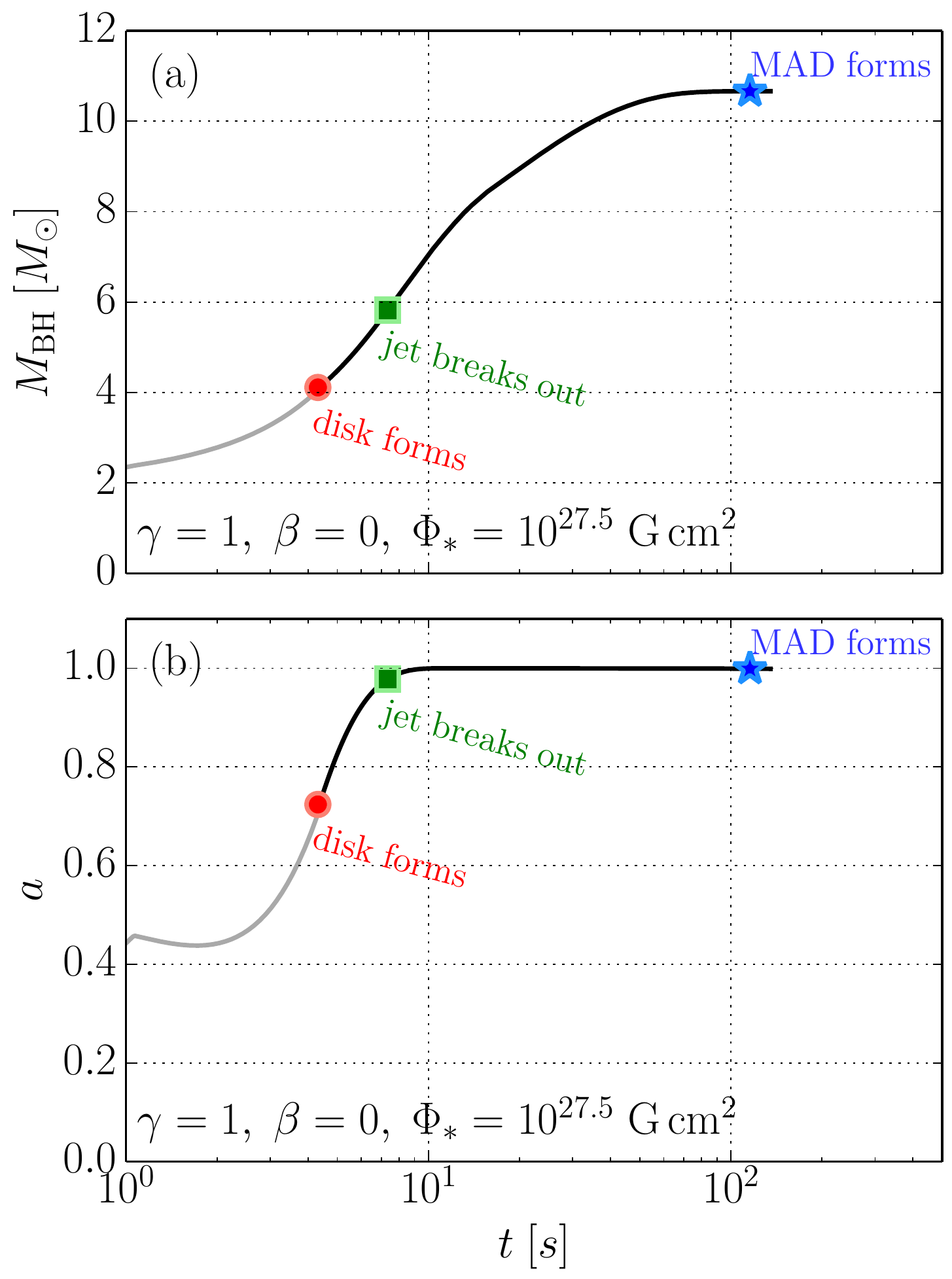}
  \end{center}
  \caption{Black hole mass $M_{\rm BH}$ (panel a) and spin $a$
    (panel b) vs.\ time $t$ since the
    core-collapse for our fiducial model. 
    Red circle indicates the time of accretion disk formation and
    green circle indicates the time of the jet breakout, which we
    associate with the gamma-ray trigger. Blue stars
    mark the time at which the magnetic flux on the BH becomes
    dynamically-important, a magnetically-arrested disk (MAD)
    forms, and the GRB ends. See Fig.~\ref{fig:mdot} for details.
  }
\label{fig:aM}
\end{figure}

\begin{figure}
\begin{center}
    \includegraphics[width=0.8\columnwidth]{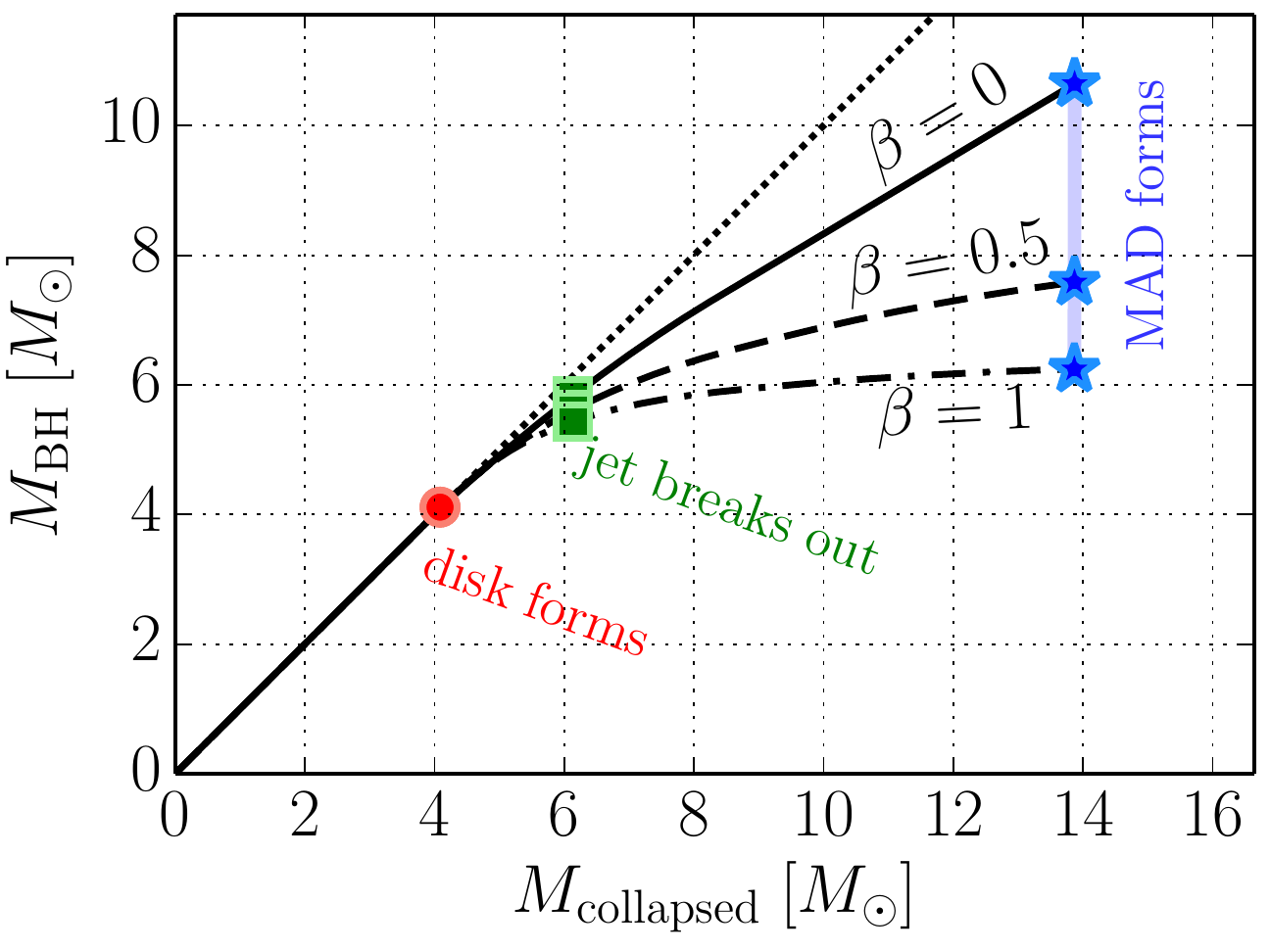}
  \end{center}
  \caption{Dependence of BH mass on the collapsed stellar mass for our fiducial
    stellar model 16TI \citep{woosley_progenitor_2006}.  The solid
    line shows the mass of the BH for our fiducial model, which does
    not include any mass-loss from
    the accretion disk ($\beta = 0$, see eq.~\ref{eq:beta}).
    The dotted line shows the
    line of equal masses. The final $M_{\rm BH}$ is lower than $M_{\rm
      collapsed}$ by about $30$\%.  This is due to the fact that during
    accretion on a rapidly spinning BH, a
    substantial part of the mass-energy is carried away by
    neutrinos, disk winds, etc.  Dashed and dash-dotted lines
    show cases with mass loss from the accretion disk:
    $\beta = 0.5$ and $1$, respectively (see eq.~\ref{eq:beta} for $\beta$
    definition). The presence of mass-loss
    further decreases the mass of the BH. The time of disk formation
    is indicated with the red circle,
  jet breakout with green square, and MAD formation with blue stars
  (which are connected with a solid light blue).}
\label{fig:Mbh}
\end{figure}

\subsection{Temporal evolution of the GRB central engine}
\label{sec:temp-evol-grb}
The collapse of the core causes the star to lose pressure support, and
the star evolves on the free-fall timescale $t_{\rm ff}\equiv r_*/v_{\rm
  ff} = (r_*^3/2GM_*)^{1/2} \sim 130$~s.  We assume that shortly after
the core collapses, a BH forms. The density profile of the star
determines the growth rate of the BH. 
The rotational profile of the collapsing star, shown in
Fig.~\ref{fig:rhoell}(b), controls which layers of the star collapse
directly into the BH. The total angular momentum $J(t)$ and mass
$M(t)$ of the layers that collapsed determine the evolution of BH spin
$a=J/Mr_gc$. When the specific angular momentum of a layer exceeds
that of a test particle orbiting at the innermost stable circular
orbit (ISCO), $l_{\rm ISCO}$, the fallback accretion hits the
centrifugal barrier, circularises and forms an accretion disk. When a
disk is present, the specific angular momentum of the material that
accretes on the black hole is set equal to $l_{\rm ISCO}$.\footnote
{We assume that the timescale for the material to reach the BH is
  dominated by the free-fall time $t_{\rm ff}$ and not by the accretion time 
  $t_{\rm acc}$. We have checked that for judicial viscosity parameter
  $\alpha=0.1$, and disk aspect ratio $H/R\sim 1$, the accretion
  timescale is nearly always shorter than the free-fall time of the
  star. In this paper, for simplicity and transparency, we ignore the
  modifications to the time it takes the gas to reach the BH due to the
  formation of an accretion disk (see
  Appendix~\ref{sec:timesc-involv-core}).  } We now discuss these
processes in more detail.

 In Fig.~\ref{fig:mdot}(a) we
show mass inflow rate $\Mdot \equiv 4\pi R^2\rho (R) v_{\rm
  ff}=4\pi R^3 \rho(R)/t_{\rm ff}$ as a function of time since core
collapse. For our fiducial model, the formation of the disk takes place at
$t_{\rm disk} \simeq 4$~seconds post core collapse and is indicated in
Fig.~\ref{fig:mdot}(a) with red circles. As seen in
Fig.~\ref{fig:mdot}(b), at this early time
the BH accretes dense, inner layers of the star, $r\sim{\rm few}\times
10^9$~cm, and the mass accretion rate is very high $\dot M\sim
M_\odot\; {\rm s}^{-1}$.\footnote{Here we equate the 
inflow rate of mass from the collapsing star to the accretion rate to
the black hole.  In effect, we ignore mass loss that can take place though, e.g.,
winds from an accretion disk. Wind mass loss is considered in
Sect. 3.3.} The accretion power, $\dot Mc^2 \sim
10^{54}\ {\rm erg\;s^{-1}}$, by many orders of magnitude exceeds the
upper range of GRB luminosity, $L_j\sim 10^{50}\ {\rm
  erg\;s^{-1}}$. This implies that, at least at early times, magnetic fields are
dynamically subdominant relative to the accretion flow (we discuss
this in Sec.~\ref{sec:bhmadestimate}).  

We assume that the formation of the centrifugal barrier and of the
disk opens up a low-density polar funnel region through which the jet emerges.
The jet is powered by BH rotation and large-scale
magnetic fields threading the BH (see Section~\ref{sec:evol-jet-lumin}
for a discussion of jet properties). The jet takes several seconds to
drill a hole through the collapsing star.  The GRB trigger takes place
at the moment the jet emerges, or ``breaks out'', of the star. For our
fiducial model this happens a few seconds after disk formation, at $t
\approx 7$~seconds after core-collapse, as indicated with green
squares in Fig.~\ref{fig:mdot}.  The trigger occurs
early in time, $\lesssim 10$~s after the core collapse, while mass
accretion rate is still very high. 
We adopt
the following expression for jet breakout time
\citep{2014arXiv1402.4142B,2014arXiv1407.0123B}:
\begin{align}
  \label{eq:ttrig}
  &t_{\rm breakout} - t_{\rm disk} 
\notag \\
  &= \left(\frac{L_j}{2\times10^{50}\ {\rm erg\; s^{-1}}}\right)^{-1/3}
   \left(\frac{M_*}{15M_\odot}\right)^{1/3} 
   \left(\frac{r_{\rm H}/a}{1.25\times 10^6\ {\rm [cm]}}\right)^{2/3}\,{\rm [s]},
\end{align}
which is the difference between the time it takes the jet head to traverse the
star and the light travel time across the star.
Here $r_{\rm H}= r_g[1+(1-a^2)^{1/2}]$ is the radius of BH event
horizon, $L_j$ is jet power (of both jets) at disk formation,
and $M_*$ the progenitor mass. Estimate~\eqref{eq:ttrig}
accounts for the fact that deep inside the star the
velocity of jet's head is non-relativistic: whereas the
most powerful jets take roughly a light crossing time to drill through
the star and break out of it (and therefore $t_{\rm breakout}-t_{\rm
  disk}\approx0$), weaker jets or
jets powered by slowly rotating BHs propagate slower and
take a longer time. As we will see below, qualitatively, 
our results are insensitive to the
particular expression for jet breakout time.

Figure~\ref{fig:aM} shows how BH mass and spin change during the core
collapse:  after the GRB trigger, BH mass changes by no more than
a factor of $2$ and the spin is practically unchanged, $a\simeq 1$.
Due to the high
value of BH spin, the gas that reaches the BH from the ISCO is strongly
gravitationally bound to the BH.  The strong binding can be achieved 
because a substantial fraction of
rest-mass energy is lost in the form of neutrinos, disk winds, etc. 
Energy loss suppresses the rate of BH growth: the black solid line in 
Fig.~\ref{fig:Mbh} shows that the final BH mass is about $30$\% less
than the mass of the collapsed star. In Sec.~\ref{sec:mass-loss-from} we discuss another
potential mechanism that can be operating in accretion flows and that can
also lead to the suppression of BH growth: mass loss from the
accretion disk. 

At late time, $t\gtrsim 50$~seconds, the BH starts to
consume the outer layers of the star (see Fig.~\ref{fig:mdot}b), in
which the density has a particularly steep radial profile (see
Fig.~\ref{fig:rhoell}a), and the time-dependence of mass accretion
rate steepens exponentially, with effective power-law scaling as steep
as $\Mdot\propto t^{-20}$ (see
Fig.~\ref{fig:mdot}a).
Thus, we
see that despite the mass accretion rate rapidly evolving in time, BH
mass and spin saturate around the start of the GRB and do not change
appreciably after that, with important consequences for jet power that
we now discuss.

\subsection{Available magnetic flux and jet power}
\label{sec:estimates}

An important factor that controls jet power is large-scale magnetic
flux threading the BH (i.e., the flux that crosses the BH event horizon).
It is conceivable that this flux is dragged to the BH by the collapsing stellar gas. We first
argue that the star can contain the necessary magnetic flux to power a GRB and
then show that the accretion disk can hold this flux on the BH
for a minute or so. We then suggest that this timescale is what sets the GRB
duration.

\subsubsection{Magnetic field strength on the hole}
\label{sec:bhestimate}

Can the strength of the magnetic flux on the BH be limited by the amount of
magnetic flux available in the progenitor star? To find this out, we will
assume that a progenitor star, similar to model 16TI of
\citet{woosley_progenitor_2006}, has a magnetic field of $B_{\rm
  surf}\sim10^4$~G at the surface, which is located at $r_*
\approx 5\times10^{10}$~cm. Since GRBs are rare events \citep{2005ApJ...619..412G},
such a strong surface magnetic field might not be uncommon among GRB progenitors.

Part of the core, with radius $r_{\rm coll}\simeq3\times10^9$~cm
(see Fig.~\ref{fig:mdot}b), directly collapses and forms a black hole
of mass $M_{\rm BH}\approx4M_\odot$ (Fig.~\ref{fig:aM}a) and
gravitational radius $r_g=GM_{\rm BH}/c^2\approx6\times10^5$~cm.
Assuming dipolar field in the star, the magnetic field strength at $r=r_{\rm
  coll}$ before the collapse is $B_{\rm coll}=(r_*/r_{\rm coll})^3B_{\rm
  surf}\approx 5\times 10^7$~G.  If the magnetic field
is frozen into the stellar envelope, the direct
collapse leads to the field strength at the BH horizon,
$B_{\rm H}\approx (r_{\rm coll}/r_g)^2 B_{\rm core}\approx
10^{15}$~G, or the magnetic flux through the BH of $\Phi_{\rm BH} \sim 5\times 10^{27}\
{\rm G\;cm^2}$. The resulting jet power,
\begin{equation}
L_{\rm BZ} \approx \frac{kf}{4\pi c}
\PhiH^2\OmegaH^2,
\label{eq:pbz}
\end{equation}
is $\sim 10^{50}\ {\rm erg\;s^{-1}}$, where $k\approx 0.05$ and $\OmegaH = ac/(2r_{\rm H})$ is the angular
frequency of BH horizon \citep{tch10a}.  We will refer to the expression for BZ power,
given by eq.~\eqref{eq:pbz} with $f=1$,
as the second-order `BZ2' approximation: it is accurate for $a\lesssim 0.95$. For our numerical models, which we
describe in Sec.~\ref{sec:evol-jet-lumin}, we will
make use of a more accurate sixth-order `BZ6' expression for jet power that is valid
for all values of spin and includes a high-order
correction factor, $f =
1+0.35\omega_{\rm H}^2-0.58\omega_{\rm H}^4$, where $\omega_{\rm H} =
a/[1+(1-a^2)^{1/2}]$ is the dimensionless rotational frequency of BH event horizon.

Thus, a simple estimate \eqref{eq:pbz} shows that the large-scale magnetic flux
contained by a progenitor star is sufficient to 
account for the power of typical long-duration GRBs, around
$10^{50}\ {\rm erg\;s^{-1}}$.
This also implies that GRBs seem to require a strongly magnetised pre-collapse core with $B\gtrsim 10^7$ G. 
Given that some white dwarfs are observed with surface magnetic fields
in excess of $\sim 10^8$~G
 \citep{2007ApJ...654..499K,2009A&A...506.1341K} such values may
be reasonable. 

\subsubsection{Is Jet Power Limited by Available Stellar Magnetic Flux
  or Mass Accretion Rate?}
\label{sec:bhmadestimate}

In Sec.~\ref{sec:bhestimate}, we found that  based on the observed energetics of GRBs, the central BH
is threaded by magnetic field of order $B_{\rm H} \sim
10^{15}$~G, or magnetic flux $\PhiBH\sim 5\times 10^{27}\ {\rm
  G\,cm^2}$. We showed
that it is conceivable that a progenitor star contains this
amount of magnetic flux.  

What keeps this magnetic flux on the BH?  If the accretion ceased, the
magnetic flux would have left the BH due to the no-hair theorem. 
Thus, it is the presence of an accretion disk via its pressure
(ram plus thermal) that keeps the magnetic flux on the BH. Immediately after the
core-collapse, the mass accretion rate is so high that the pressure of
the disk easily overpowers the outward magnetic pressure. However, as
$\Mdot$ drops, eventually the disk pressure becomes too weak to hold
the entirety of magnetic flux on the BH. This happens at a critical
mass accretion rate at which accretion and jet power are comparable
(\citealt{tch11,tch12proc,2014MNRAS.437.2744T}, Tchekhovskoy, McKinney
and Nemmen, in preparation),
\begin{equation}
  \label{eq:mmadsimple}
  \Mdot_{\rm MAD} c^2 \approx  L_j 
    \left(\frac{a}{1}\right)^{-2}
    \left(\frac{h/r}{0.2}\right)^{-1}.
\end{equation}
Here $h/r=0.2h_{0.2}$ is half-thickness of the inner regions of the
accretion disk (the disk is kept relatively thin by neutrino cooling; \citealt{chen_neutrino_cooled_2007}).
In terms of magnetic flux threading the BH,
\begin{equation}
  \label{eq:mdotmad}
  \Mdot_{\rm MAD} = 10^{-4} M_\odot\,{\rm s^{-1}}
  \left(\frac{\PhiBH}{5\times 10^{27}\ \Gcmsq}\right)^2
  \left(\frac{M_{\rm BH}}{10M_\odot}\right)^{-2}
  \left(\frac{h/r}{0.2}\right)^{-1}
  \chi^{-2},
\end{equation}
where 
$\chi=1.4(1-0.38\omega_{\rm H}) \sim 1$ is a high-order correction.

Typically, collapsars have $\Mdot \gtrsim 0.1 M_\odot\ {\rm s^{-1}}\gg
\Mdot_{\rm MAD}\sim 10^{-4}M_\odot\ {\rm s^{-1}}$.  Thus, the accretion rate onto a newly formed BH is
more than sufficient to confine the magnetic flux within the black
hole vicinity. And, as long as $\Mdot \ge \Mdot_{\rm MAD}$, BH
magnetic flux $\PhiBH$ is determined by the available stellar magnetic flux
and not by mass accretion rate, even though
$\Mdot$ drops fast with time (see e.g.~Fig.~\ref{fig:mdot}). Therefore, after the jet emerges
through the stellar surface $\sim 10$~s post core collapse, it
powers a GRB of roughly a constant luminosity (see eq.~\ref{eq:pbz}),
which is illustrated by horizontal segments of the light curves in Fig.~\ref{fig:sketch}.

After a minute or so the situation qualitatively changes:
$\Mdot$ drops below $\Mdot_{\rm MAD}$, the magnetic flux on the hole
becomes dy\-na\-mi\-ca\-l\-ly-important, and parts of the
flux, which used to thread the BH, diffuse
out.  The remaining BH magnetic flux obstructs gas infall and
leads to a \emph{magnetically-arrested disk} (MAD,
\citealt{nia03,igu03,tch11}, see also
\citealt{1974Ap&SS..28...45B,1976Ap&SS..42..401B}). The
remaining BH magnetic flux is
determined by the instantaneous $\Mdot$ via eq.~\eqref{eq:mdotmad},
\begin{equation}
  \label{eq:phibhmad}
  \Phi_{\rm BH,MAD} = 5\times 10^{27}\ {\rm G\;cm^2} 
             \left(\frac{\Mdot}{10^{-4}M_\odot\,{\rm s^{-1}}}\right)^{1/2}
             \left(\frac{M_{\rm BH}}{10M_\odot}\right)
             \left(\frac{h/r}{0.2}\right)^{1/2}\chi,
\end{equation}
and produces jets of energy efficiency, or dimensionless power, given by eq.~\eqref{eq:mmadsimple}:
\begin{equation}
\eta_{\rm MAD} \equiv \frac{L_{j,\rm MAD}}{\Mdot c^2}\simeq a^2 h_{0.2},
\label{eq:pmad}
\end{equation}
i.e., there is a linear relation of $\Mdot$ and jet
power $L_{j,\rm MAD}$ in the MAD regime. Since $\Mdot$ rapidly decreases in time, after the MAD onset
we expect the jet power $L_{j,\rm MAD}$ to do so as well. Note that in our numerical models
described in Sec.~\ref{sec:evol-jet-lumin}, we will use a more accurate
expression for jet energy efficiency,
  \begin{equation}
\eta_{\rm MAD}=3 \omega_{\rm
    H}^2 (1-0.38 \omega_{\rm H})^2(1+0.35\omega_{\rm
    H}-0.58\omega_{\rm H}^2)h_{0.2}, \label{eq:etamad}
\end{equation}
which we obtain by combining the
high-order accurate versions of eqs.~\eqref{eq:pbz} and
\eqref{eq:phibhmad}.  Equation~\eqref{eq:pmad} is a quadratic
approximation to the dependence \eqref{eq:etamad}.

\section{Evolution of jet luminosity during prompt phase}
\label{sec:evol-jet-lumin}

\subsection{Magnetic flux and its ditribution in the progenitor}
\label{sec:mag-flux}

\begin{figure}
\begin{center}
    \includegraphics[width=0.9\columnwidth]{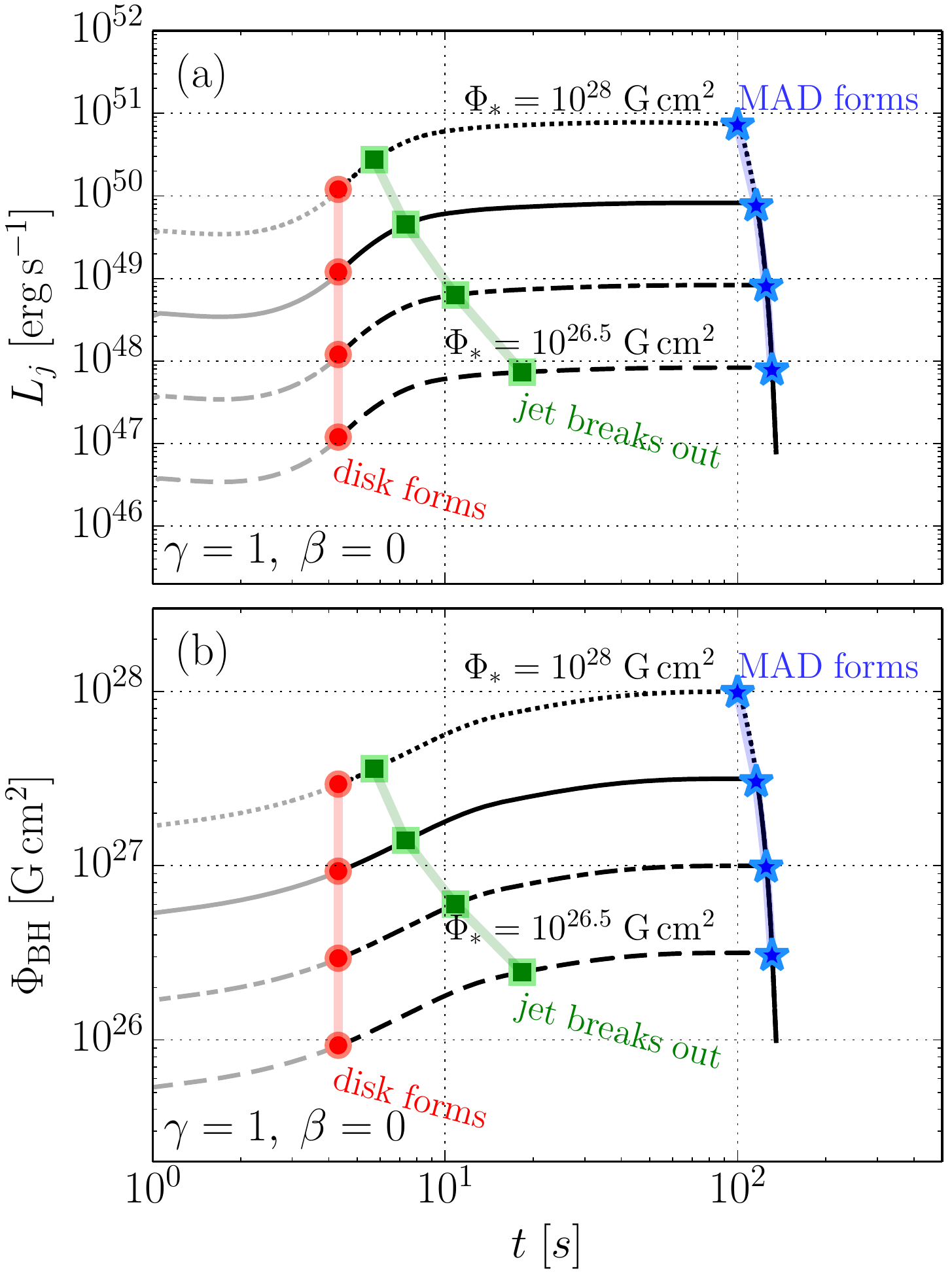}
  \end{center}
  \caption{Panels (a) and (b) show, respectively, jet power and
    magnetic flux vs. time, for different values of magnetic fluxes in
    the progenitor star. Red circles indicate the time of disk
    formation, green squares jet breakout (=GRB trigger), and blue stars the
    formation of the MAD (=end of the GRB).}
\label{fig:pjet}
\end{figure}

\begin{figure}
\begin{center}
    \includegraphics[width=0.9\columnwidth]{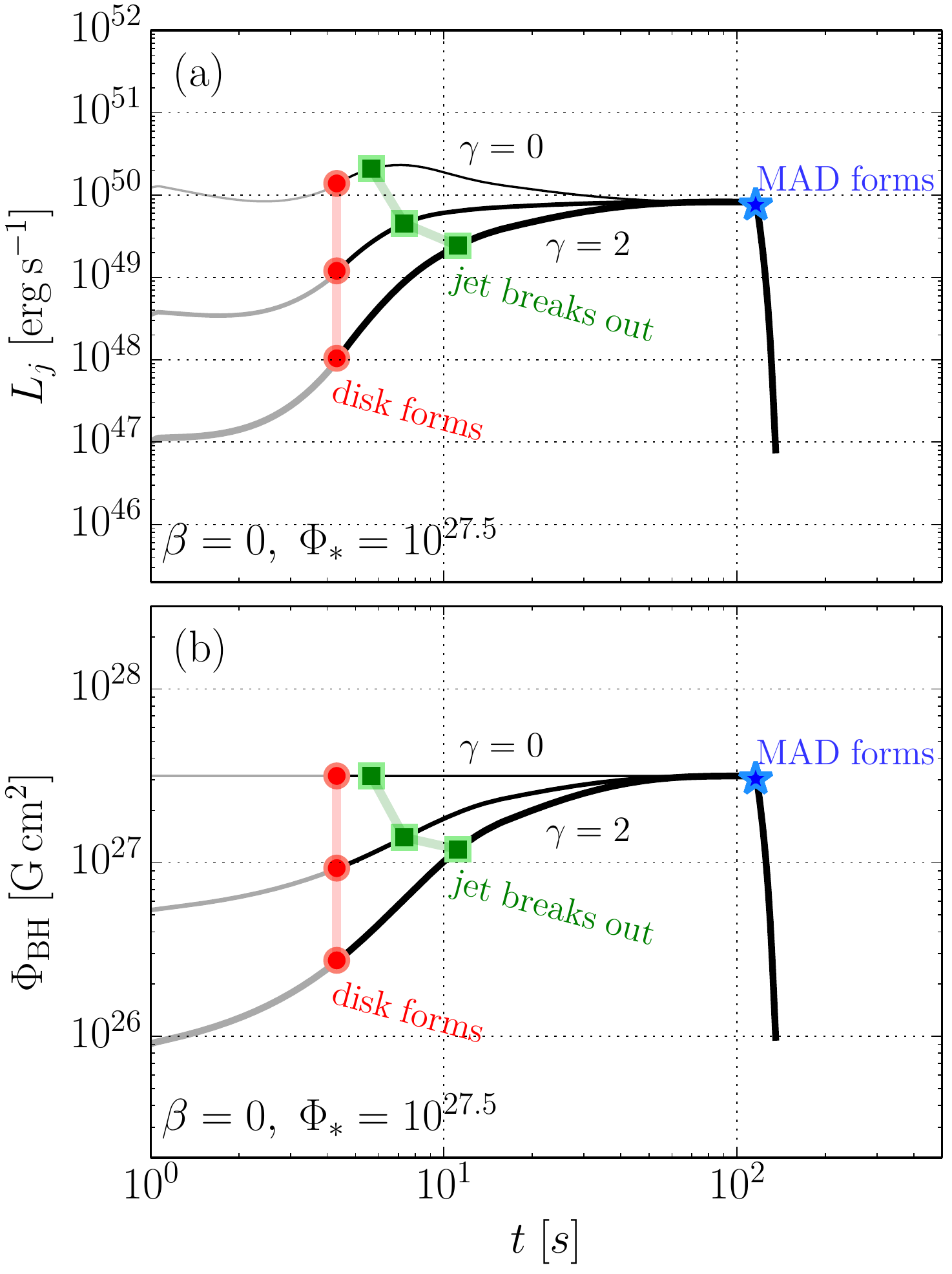}
  \end{center}
  \caption{Panels (a) and (b) show, respectively, jet power and magnetic flux
    vs. time, for different distributions of magnetic flux in the
    progenitor star: $\gamma = 0$ (thin), $1$ (medium-thickness), $2$ (thick
    solid line). Connected red circles show the times of disk
    formation, green squares
    the times of jet breakout (=GRB trigger), and blue stars the times of MAD
    formation (=end of GRB).    Variations in spatial
    distribution of magnetic flux lead to at most variations by factors
    of a few in GRB luminosity for the different models we considered.}
\label{fig:pjetphibhalpha}
\end{figure}

We have argued that GRB jets are powered by the rotational energy of
the central BH extracted by large-scale magnetic flux 
(see eq.~\ref{eq:pbz}). Within this framework, in
order to get a handle on what determines the luminosity of a GRB, we
need to have a better understanding of the processes that control the
accumulation of large-scale vertical magnetic flux on the BH.

The growth of the magnetic flux on the BH depends on the very
uncertain magnetic field configuration in the progenitor star. However, since the
flux is brought in with the accreting gas, it is reasonable to
parametrize it to scale with the collapsed mass of the star:
\begin{equation}
\Phi_{\rm BH} (t) = \Phi_* \left(\frac{M_{\rm collapsed}}{M_*}\right)^\gamma,
\label{eq:PhivsM}
\end{equation}
where $\gamma\sim 1$ parametrizes the radial
distribution of magnetic flux inside of the star and $\Phi_*$ 
the total magnetic flux threading the star.
If we now combine eq.~\eqref{eq:PhivsM} with eq.~\eqref{eq:pbz}, we
obtain:
\begin{equation}
L_{\rm BZ} \propto \frac{a^2 \PhiBH^2}{\MBH^2} \propto a^2\MBH^{2(\gamma-1)}, 
\label{eq:lbzalpha}
\end{equation}
where we loosely approximated $M_{\rm BH} \approx M_{\rm collapsed}$.  Since
BH mass changes by a factor of few in the course of GRB (see Fig.~\ref{fig:aM}a),
eq.~\eqref{eq:lbzalpha} suggests that the power of the jets changes by
the same (small) factor of a few, and this might account for the
apparent constancy of GRB prompt emission we set out to explain. We now explore this
possibility in detail.
 
In Fig.~\ref{fig:pjet}(a),(b) we plot the time-dependence of jet power, or luminosity $L_j$, and the magnetic flux $\PhiBH$
threading the BH, for our fiducial scenario, $\gamma = 1$. We consider
different values of progenitor magnetic flux strength,
$\Phi_*=10^{26.5}, 10^{27}, 10^{27.5}, 10^{28}$~G~cm$^2$, as labeled
on the figure and shown with black lines of different types.
The behaviour of jet power is very similar, for the entire range of the magnetic
flux strength in the progenitor star. After the trigger, which we
indicate with the green squares, and until the MAD formation, which we
show with blue stars, both magnetic flux and jet power change by about
a factor of $2$. During this time the strength of magnetic flux on the
BH, shown in Fig.~\ref{fig:pjet}(b), is determined by the magnetic flux supply
in the progenitor star and not by mass accretion rate. Note that for our fiducial model, with $\gamma
= 1$, by eq.~\eqref{eq:lbzalpha} we would expect no dependence of jet power on $M_{\rm BH}$ (and
time) at
all. As we will see later, there is actually some dependence, caused
by the differences between $\MBH$ and $M_{\rm collapsed}$ (see
Fig.~\ref{fig:Mbh}). However, for
the entire range we considered, $0\le\gamma\le2$, the variation of $L_j$
during the GRB is rather weak. 
After
about $100$ seconds, a MAD forms, and the luminosity drops fast since
the disk cannot hold onto the BH magnetic flux any more and the 
GRB turns off. We discuss this in detail in Sec.~\ref{sec:steep-decline-phase}.

Note that variation by a factor
of $\sim$~30 of the flux $\Phi_*$ available in the progenitor star,
reproduces the full range of observed GRB
luminosities ($L_j\propto \Phi_*^2$),
from the weakest to the most powerful ones. This range
corresponds to surface (core) field strength of the progenitor ranging from
$10^3$~G to $3\times 10^4$~G ($3\times 10^6-10^8$~G). Indeed such 
a factor of $\sim$~30 is well within the observed range of variation 
in the surface field strength of massive stars. The field
strength is the only property of the progenitor star that we
are aware of, that
can plausibly give the large variety of GRB energetics, as observed (see Sec.~\ref{sec:effect-prec-model}).

Figure~\ref{fig:pjetphibhalpha}(a),(b) explores the effect of different
radial distributions of $\Phi_*$ in the progenitor star. For this, we
plot the time-dependence of $L_j$ and $\PhiBH$ for 3 different
values of the $\gamma$ parameter (defined in eq.~\ref{eq:PhivsM}):
$\gamma = 0,1,2$. The variation in $\gamma$ affects
neither the time of GRB turnoff nor the characteristic power of the
GRB, but it does change the early-time trend of jet power dependence
on time: for $\gamma=0$, shown with the thin solid line, jet power
decreases with time, before levelling off at a constant value. 
Thus, even though the magnetic flux on the BH is constant in this case
(eq.~\ref{eq:PhivsM}),
jet power changes: this is because in addition to BH magnetic flux,
jet power also depends on the BH mass, whose growth in time
leads to the decrease of jet power (see eq.~\ref{eq:pbz}).  For
the other two cases, $\gamma =1$ and $2$, both $\PhiBH$ and $L_j$
increase with time before levelling off at constant values. 
We conclude that apart from
early time trends, the light curves are not sensitive to the choice of
$\gamma$.

\subsection{Steep Decline Phase}
\label{sec:steep-decline-phase}
Steep decline stage is routinely seen by \emph{Swift}/XRT as a rapid
turn-off of GRB luminosity at the end of the GRB, as $L_j\propto t^{-3..-5}$. In our model, the
steep decline stage starts at the formation of the MAD. In this steep
decline stage, we are likely seeing the turn-off of the central engine:
the decrease in mass accretion rate below $\Mdot_{\rm MAD}$ (see
eq.~\ref{eq:mdotmad}) leads to the MAD state, in which the jet
power tracks the rapidly declining $\Mdot$. Thus, we can conclude that the
observed steep decline is the natural result of the central BH
accreting the outer layers of the star. In fact, we find that the jet
power drops extremely steeply, $L_j \propto \Mdot \propto t^{-20}$ or
so.  The resulting gamma-ray luminosity, $L_\gamma$, will experience a
shallower decline, depending on the size of the emitting region
\citep{2000ApJ...541L..51K}. Thus, in order to explain the observed decline, $L_\gamma \propto
t^{-3..-5}$, the jet has to turn-off \emph{at least} as steeply as the
bursts are observed to decline, which is consistent with our findings.

\begin{figure}
\begin{center}
    \includegraphics[width=1.05\columnwidth]{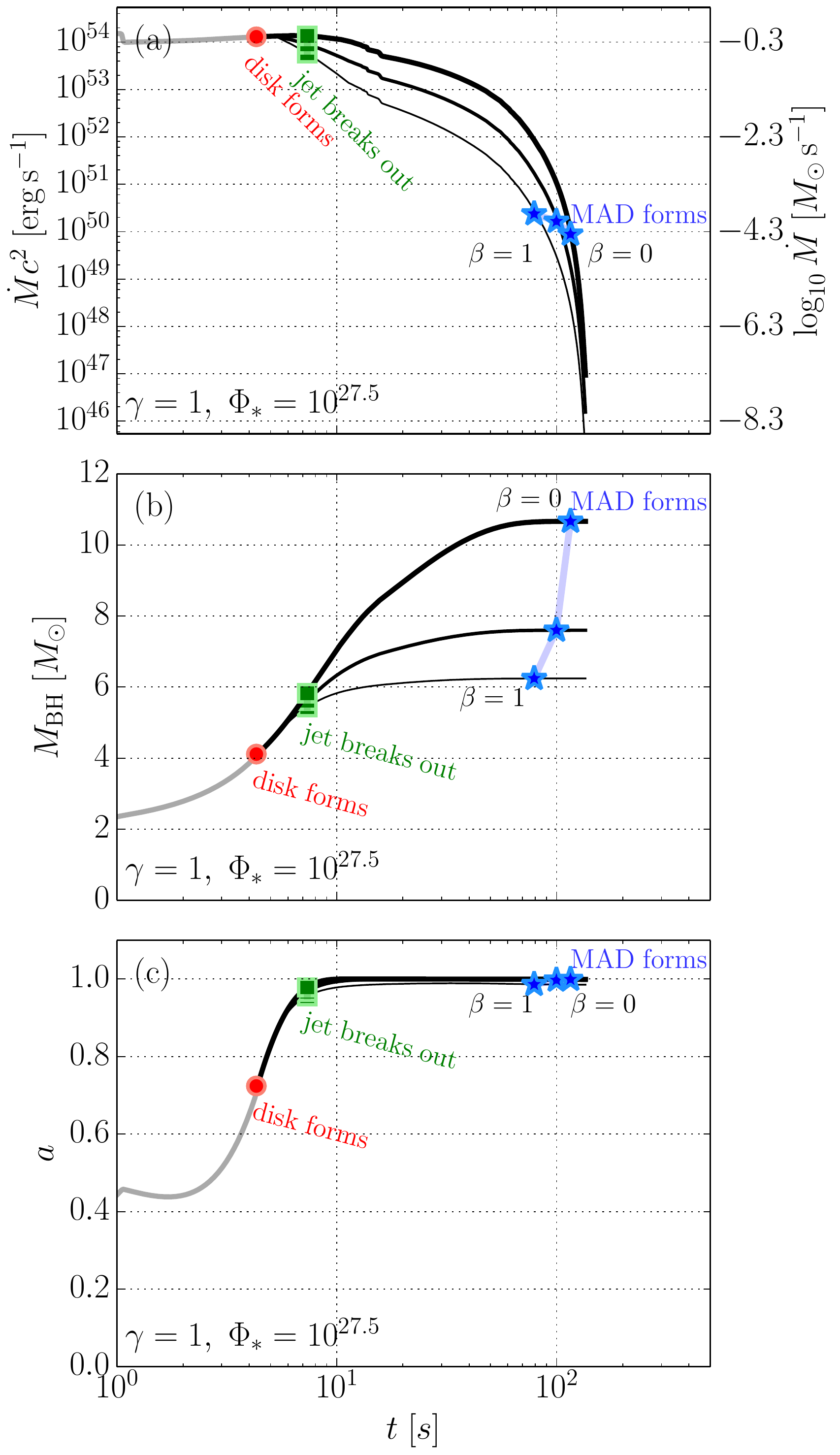}
  \end{center}
  \caption{Mass accretion rate $\dot M$, BH mass $M_{\rm BH}$, and BH
    spin, $a$ as a function of time for the case when we explore the
    effect of mass loss in a wind from an accretion disk. The
    mass-loss is modelled through suppression of $\dot M$ versus
    radius, $\dot M = (10r_g/r_{\rm D})^\beta \dot M_{\rm D}$, where
    $r_{\rm D}$ is the outer radius of the accretion disk. We explore
    three different $\beta$ values: $0$, $0.5$, and $1$. }
\label{fig:massloss1}
\end{figure}

What sets the duration of the GRB in our model? There are at
least two
parameters: the strength of the magnetic flux threading the star,
$\Phi_*$, and the density distribution in the outer layers of the
star, which sets the late-time dependence of $\Mdot$. Since the
strength of $\Phi_*$ can be inferred from GRB jet luminosity via
eq.~\eqref{eq:pbz}, the end of the GRB signals the time at which
$\Mdot = \Mdot_{\rm MAD}$ (see eq.~\ref{eq:mdotmad}), i.e., the
accretion flow onto the BH enters the MAD regime. Since MADs around
rapidly spinning BHs give $L_j \approx \Mdot c^2$ (see eq.~\eqref{eq:mmadsimple}), we can estimate
mass accretion rate at the end of the GRB as
\begin{equation}
  \label{eq:mdotlate}
  \Mdot_{\rm turnoff} = \frac{L_j}{c^2} = 0.6\times 10^{-4} M_\odot\,{\rm s^{-1}} \frac{L_j}{10^{50}{\rm erg\, s^{-1}}}.
\end{equation}

Since, according to eq.~\eqref{eq:mdotmad}, we have $\Mdot_{\rm
  MAD}\propto \PhiBH^{1/2}$, stronger BH magnetic flux translates into a
higher mass accretion rate at GRB turn-off. Therefore, for larger
$\PhiBH$ and all else being equal, we would expect the GRB to be more
powerful and last for a shorter time. However, this effect is
diluted by the fact that more powerful jets breakout faster and thus
we start seeing GRB earlier (see eq.~\ref{eq:ttrig}). 
We will also see later (Sec.~\ref{sec:effect-stell-rotat}) that changes in stellar angular momentum
profile strongly affect the jet breakout time and the GRB duration.

\subsection{Mass loss from the disk}
\label{sec:mass-loss-from}

So far we have assumed that all the mass from the collapsing star
makes it to the black hole. In reality, stellar material settles into
a disk with an outer size $r_{\rm D}$, which is set by the value of
the specific angular momentum $\ell$ of the stellar material, $r_{\rm D}
\approx \ell^2/GM$, where $M$ is the enclosed mass.  For outer layers of the star, $r_{\rm D}$ can be
much larger than $r_{\rm ISCO}$, and substantial mass loss can take
place between the two radii \citep{1999MNRAS.303L...1B,2012MNRAS.423.3083M,2012MNRAS.426.3241N,2013MNRAS.436.3856S,2014arXiv1407.4421S} 

We model the effect of mass loss in a standard way, assuming that
winds take away a fraction of locally accreting mass, i.e.,
we parametrize $\Mdot$ at the BH as \citep[e.g.,][]{1999MNRAS.303L...1B}
\begin{equation}
  \label{eq:beta}
  \Mdot = \left(\frac{r_0}{r_{\rm D}}\right)^\beta\Mdot(r_{\rm D}),
\end{equation}
where $r_0 = 10r_g$ is a characteristic radius at which the disk wind,
which carries away mass from the disk, starts. The value of $\beta$
controls the strength of mass loss: $\beta = 0$ means no mass loss,
$\beta = 1$ means very strong mass loss. GR numerical simulations
suggest $\beta \approx 0.5$ \citep{2012MNRAS.423.3083M,2012MNRAS.426.3241N}.

\begin{figure}
\begin{center}
    \includegraphics[width=0.9\columnwidth]{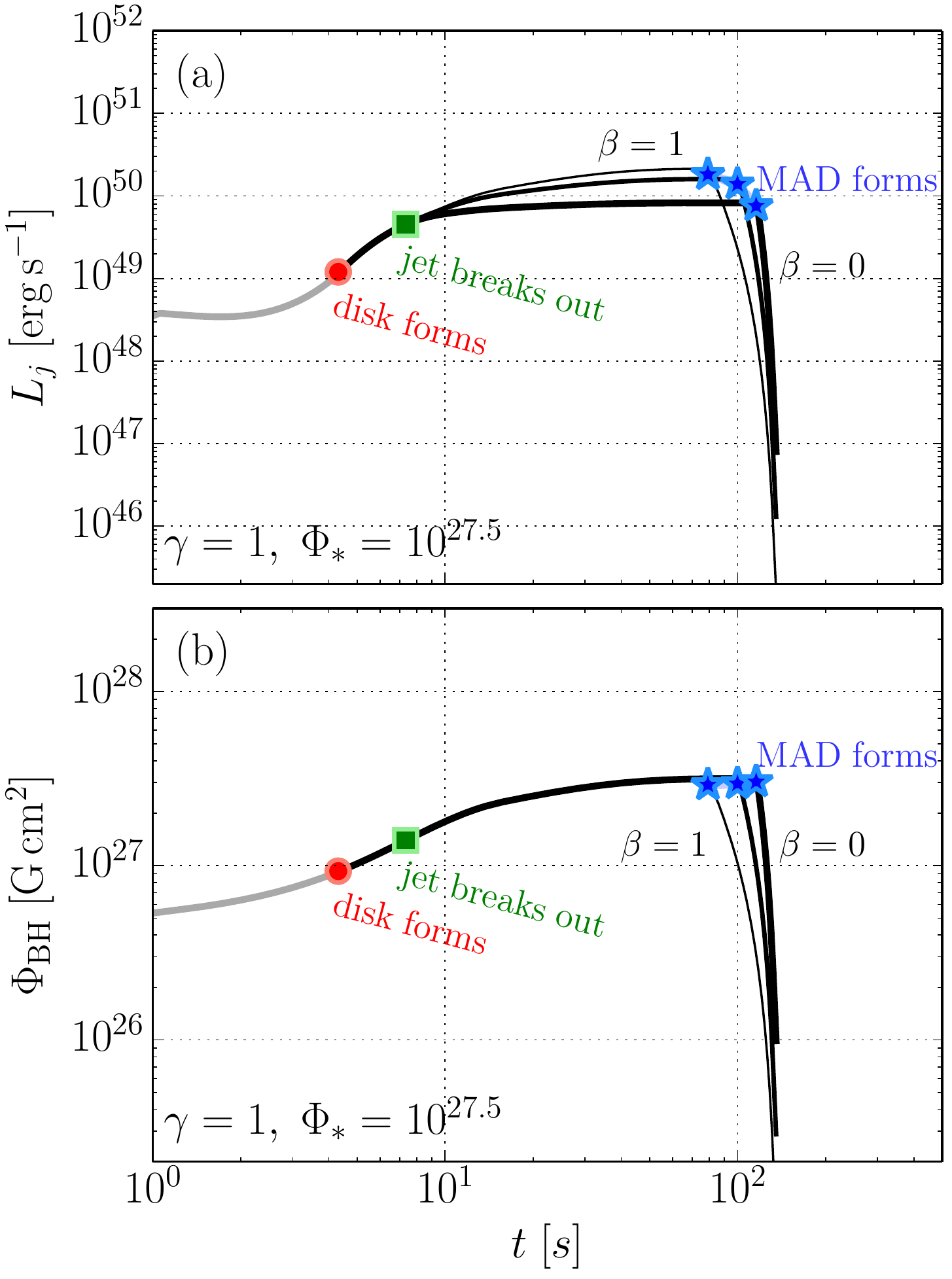}
  \end{center}
  \caption{Similar to Fig.~\ref{fig:massloss1}, but for $\Phi_{\rm
      BH}$ and $L_j$. The mass loss does not
    qualitatively affect the temporal dependence of jet
    power. However, it does change the duration of the prompt
    emission: the larger the mass loss, the earlier the MAD onset, the
  shorter the GRB.}
\label{fig:massloss2}
\end{figure}

We show the effects of disk mass-loss in Figures~\ref{fig:massloss1}
and \ref{fig:massloss2} for 3 different values of mass-loss
parameter: $\beta =
0, 0.5$, and $1$. Figure~\ref{fig:massloss1}(a) shows that the presence
of mass-loss suppresses BH mass accretion rate: the thick solid line,
which corresponds to the case $\beta = 0$ and does not include any
mass loss (see eq.~\ref{eq:beta}), lies above the other two lines for
$\beta=0.5$ and $1$, both of which include mass-loss. The suppression
of $\Mdot$ for $\beta > 0$ leads to the suppression of BH mass growth: as is
clear from Fig.~\ref{fig:massloss1}(a),(b), the
higher the value of $\beta$, the stronger the mass loss, the smaller
the BH mass. However, the plausible range of $\beta$ we considered
results in only a modest change in $\MBH$, by at most a
factor of $2$.
Figure \ref{fig:massloss1}(c) shows that mass loss does not strongly
affect BH spin, either: its value levels off at $a\simeq1$ soon after
jet breakout for the entire range of $\beta$ considered. 
This is because by the time mass-loss becomes
substantial (i.e., $r_{\rm D} \gg r_0$), 
the BH was already spun up to a near-maximum spin,
$a\gtrsim0.9$. 

\begin{figure}
\begin{center}
    \includegraphics[width=0.9\columnwidth]{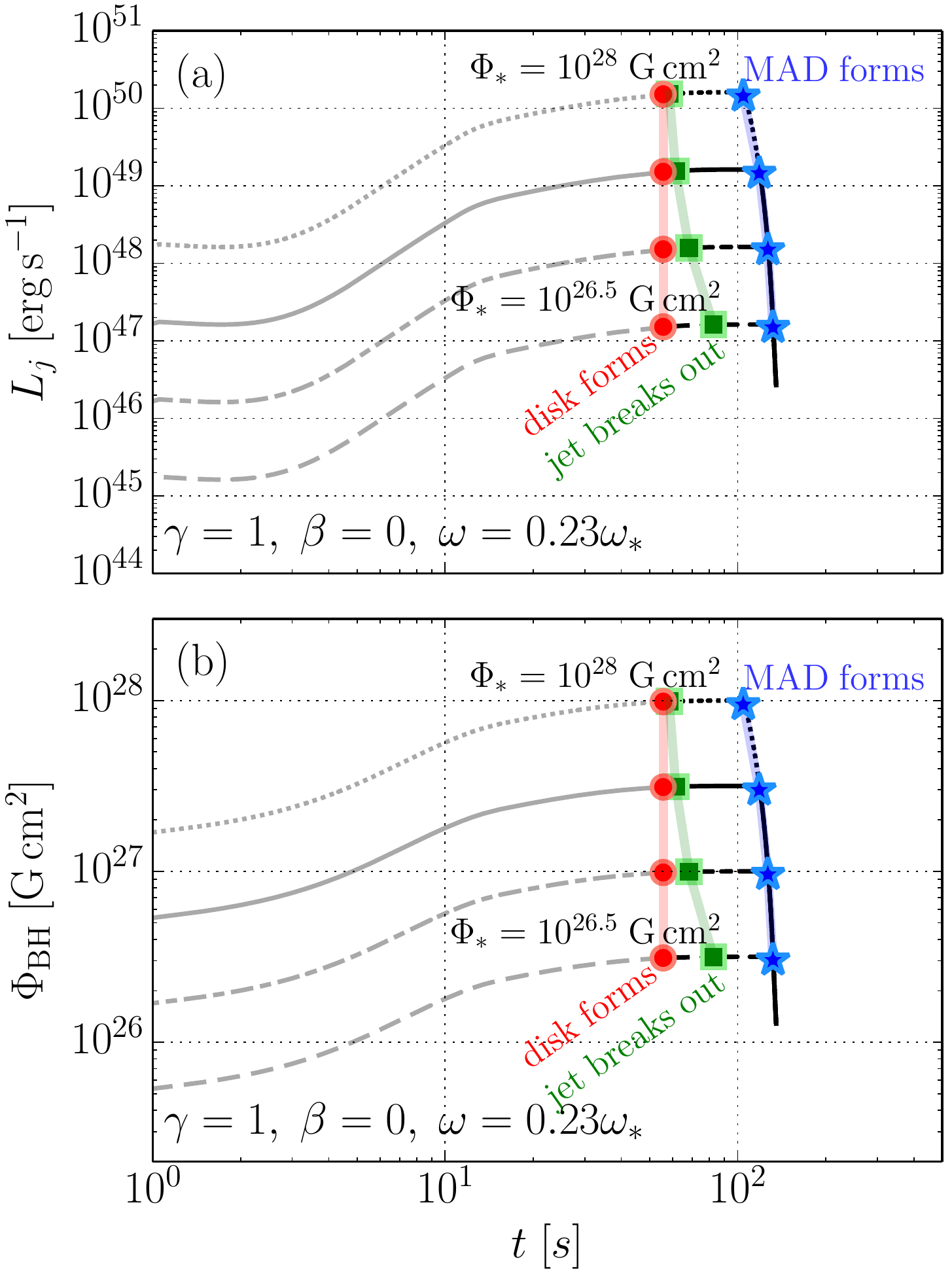}
  \end{center}
  \caption{Jet power and magnetic flux vs.\ time, for
    slower stellar rotation than in our fiducial model
    ($\omega=0.23\omega_*$) and different values of magnetic flux in
    the progenitor star, $\Phi_*$. Comparison to Fig.~\ref{fig:pjet}
    shows that slower rotation does not noticeably affect the power of
    the jets. However, it does lead to a later time of debris
    circularisation and disk formation (shown with red circles) and
    jet breakout (shown with green squares that are barely visible
    under the red circles). Since the time of MAD formation (shown
    with blue stars) stays the same, slower rotation of a
    progenitor leads to shorter GRBs.}
\label{fig:pjetslowrot}
\end{figure}

\begin{figure}
\begin{center}
    \includegraphics[width=1.05\columnwidth]{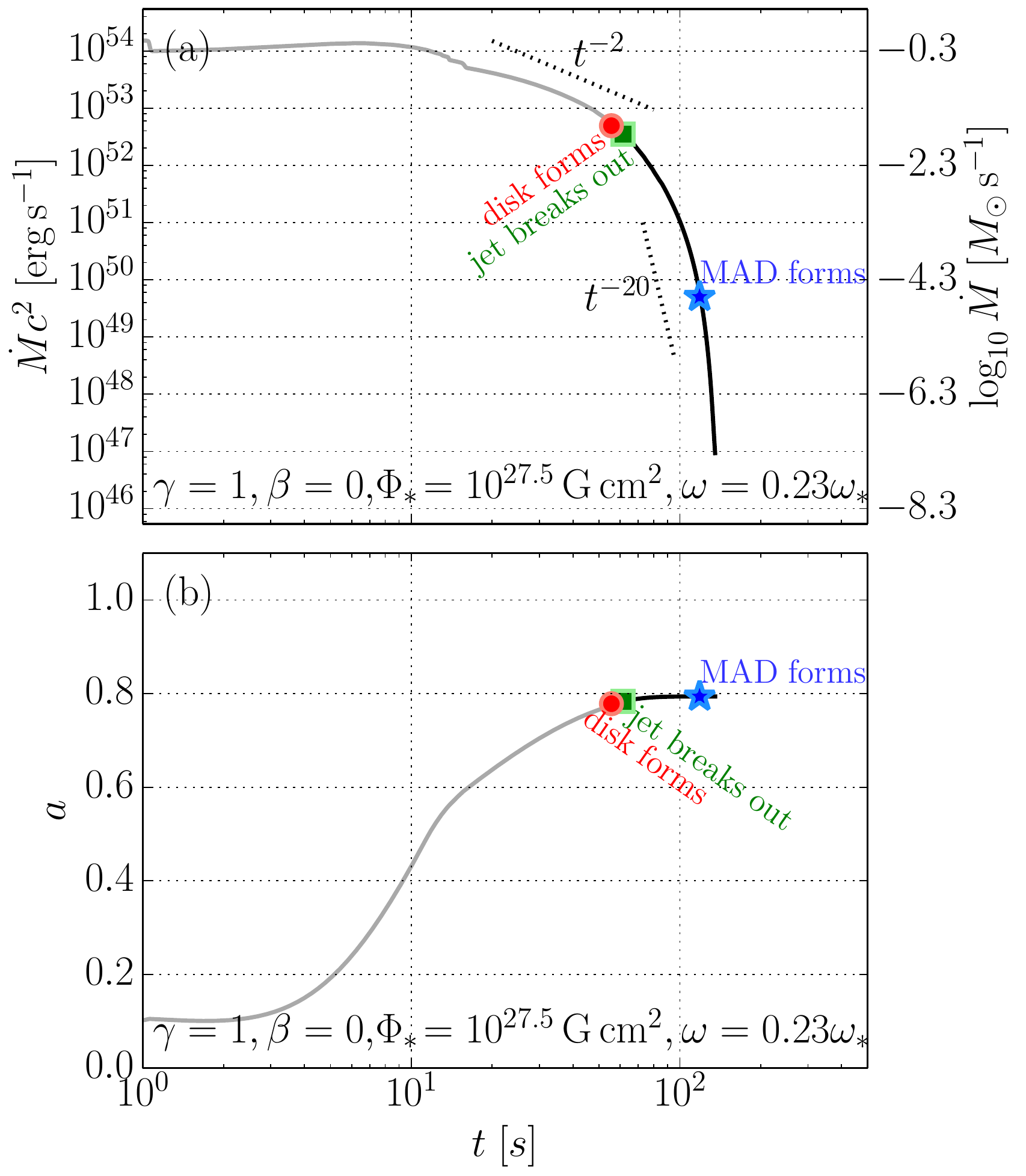}
  \end{center}
  \caption{{\bf [Panel (a)]:} The slower rotation
    of the progenitor star ($23\%$ of our fiducial model), does not
    lead to any changes in mass accretion rate: $\Mdot$ is the same as
    in our fiducial model, see Fig.~\ref{fig:mdot}(a). {\bf [Panel
      (b)]:} Despite the
    progenitor star has only $23\%$ of the angular momentum of
    our fiducial model, the collapse leads to a rapidly spinning
    BH, with $a\simeq0.8$. The rather high spin leads to BH power
    similar to our fiducial model.
    The primary effect of the slower rotation is the delay of 
    disk and jet formation, which results in a shorter GRB.}
\label{fig:mdotslowrot}
\end{figure}

\begin{figure}
\begin{center}
    \includegraphics[width=1.05\columnwidth]{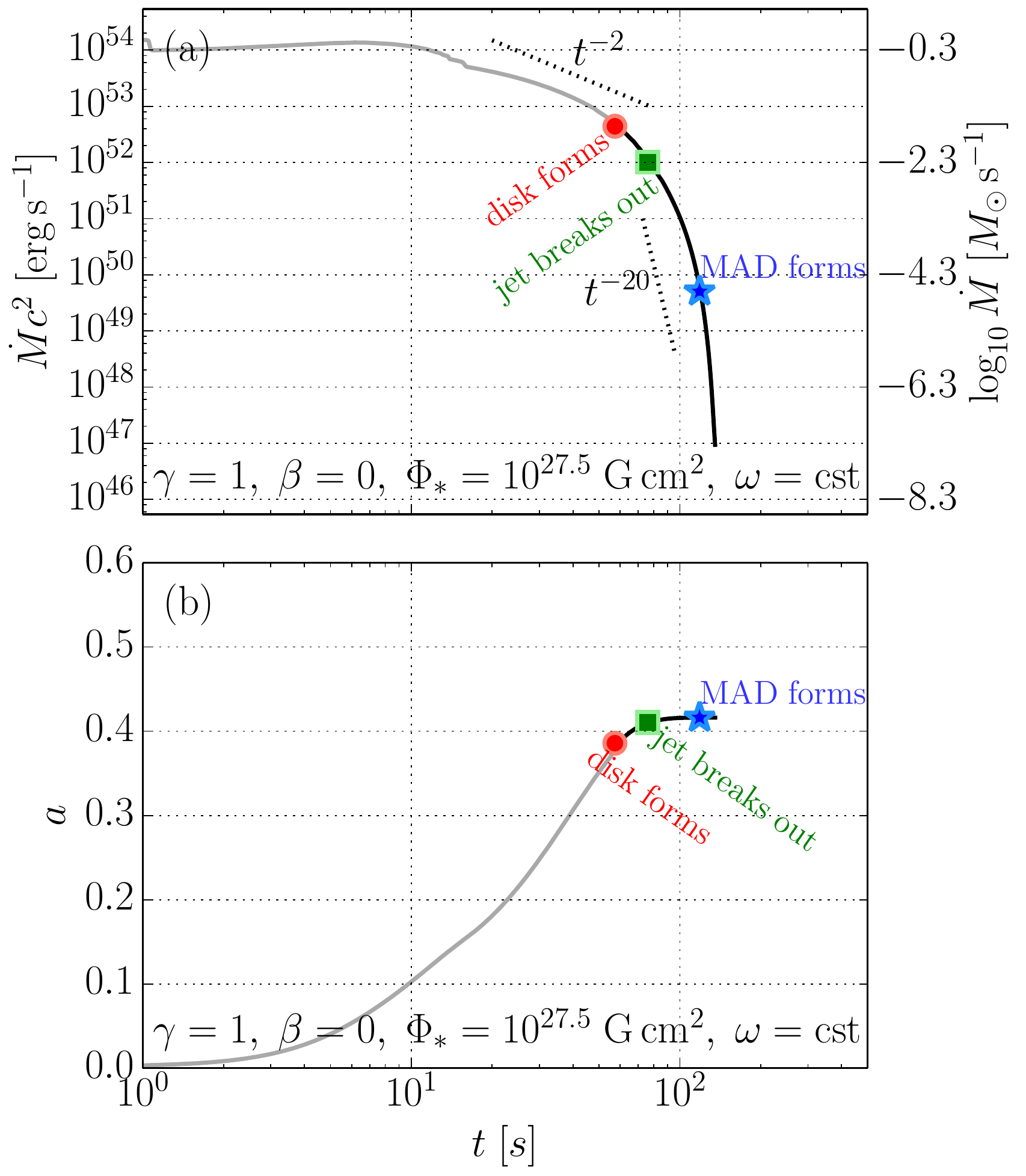}
  \end{center}
  \caption{GRB properties vs time for a model with a \emph{solid-body
      rotation} profile in the progenitor star, with surface rotation
    speed of $200$~km~s$^{-1}$, which is a characteristic rotational
    speed for O-stars. {\bf [Panel (a)]:} Time-dependence of $\Mdot$
    for a progenitor star with a solid-body rotation
    profile. $\Mdot(t)$ is the same as in our fiducial model, but disk
    formation and jet breakout happen at a later time due to the lower
    angular momentum of the progenitor star (compared to
    Fig.~\ref{fig:mdot}).  {\bf [Panel (b)]:} BH spin, $a$, vs
    time. Because for solid-body rotation $\ell\propto r^2$, most of
    the angular momentum resides in the outer layers of the star and
    reaches the BH late in the course of the core-collapse; in fact,
    direct collapse into BH of the core leads to an essentially
    non-rotating BH: $a\approx3\times10^{-2}$ at
    $t\approx1$~s. However, by the time the accreting material hits
    the centrifugal barrier and circularises, $t\approx 60$~s, BH spin
    reaches a respectably high value, $a\approx 0.4$.}
\label{fig:mdotsolidbodyrot}
\end{figure}

\begin{figure}
\begin{center}
    \includegraphics[width=0.9\columnwidth]{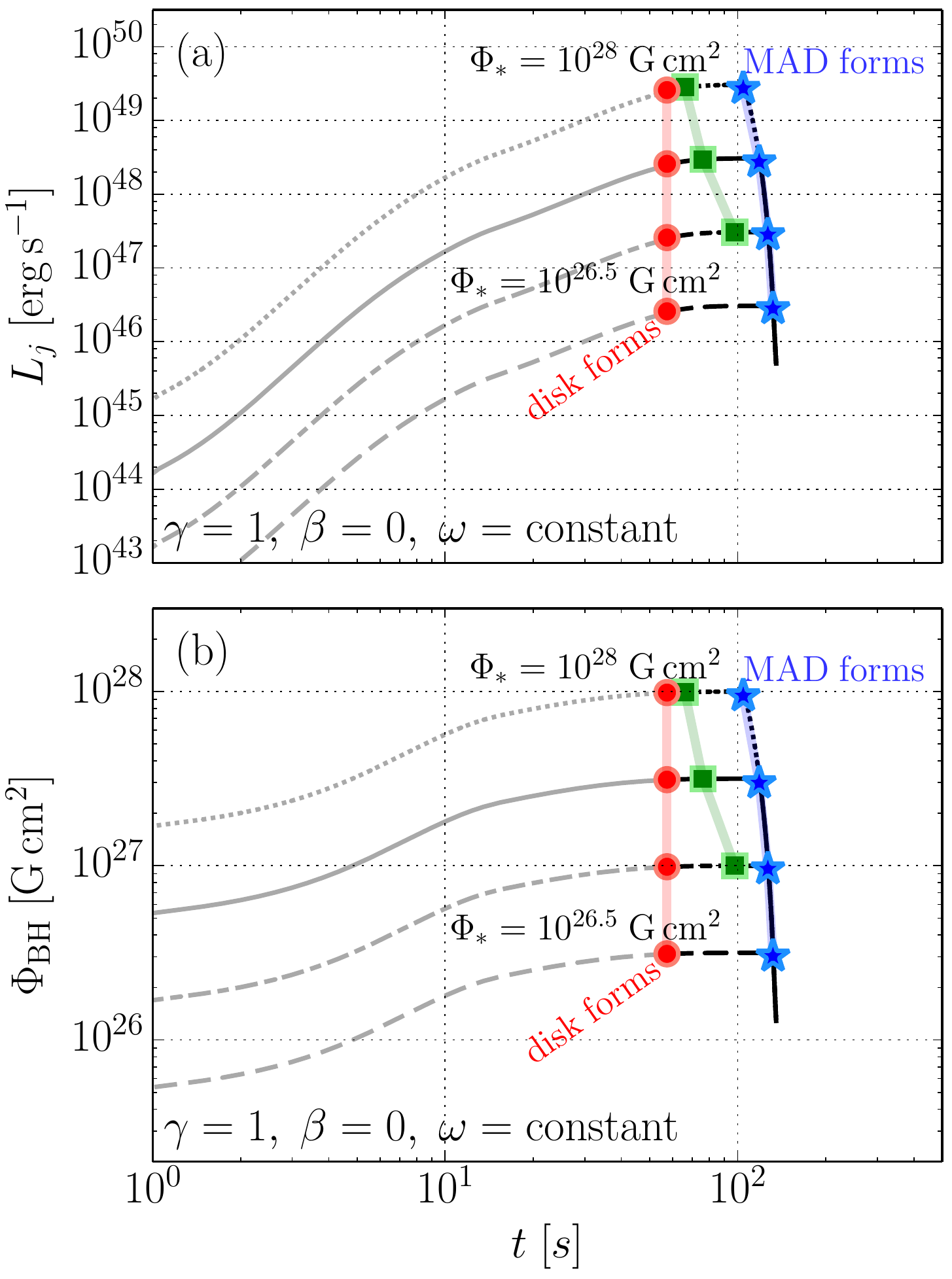}
  \end{center}
  \caption{GRB properties vs time for a model with a \emph{solid-body
      rotation} profile in the progenitor star, with surface rotation
    speed of $200$~km~s$^{-1}$, which is a characteristic rotational
    speed for O-stars. {\bf [Panel (a)]:} GRB luminosity vs time
    for different values of stellar magnetic flux, as labeled (see
    also Fig.~\ref{fig:pjet}). The relatively modest solid-body
    rotation leads to a delayed formation of disk and jet: around
    $60$~s post-collapse. Despite this delay, the lightcurve follows
    the standard GRB template, with a constant-luminosity plateau
    lasting about $40$~s (shorter for weaker magnetic flux) and
    followed by an abrupt turn-off. For the weakest magnetic flux,
    $\Phi_* = 10^{26.5}$~G~cm$^2$, the jet is so weak that it does not
    make it out of the star before the accretion turns off. {\bf
      [Panel (b)]:} BH magnetic flux $\PhiBH$ as function of time, for
    different values of total stellar magnetic flux, $\Phi_*$. The
    magnetic flux time-dependence is identical to that in our fiducial
    model shown in Fig.~\ref{fig:pjet}(b).   Thus, the primary
    effect of the solid-body rotation is the delayed time of disk and
    jet formation and shortening of the GRB duration.}
\label{fig:pjetsolidbodyrot}
\end{figure}

Figure~\ref{fig:massloss2}(a)
shows that jet power, $L_j\propto \MBH^{-2}$, is slightly increased
for $\beta > 0$, due to the
suppression of BH mass growth. Figure~\ref{fig:massloss2}(b) shows that mass loss does not change the
amount of magnetic flux brought to the BH, i.e., the three solid
lines, each for a different value of the mass-loss parameter $\beta$, are
on top of each other. This is to be expected, since in our model the
time-dependence of $\PhiBH$ is determined by the initial distribution
of magnetic flux in the progenitor star and therefore depends neither on
mass-loss nor the choice of $\beta$ (see eq.~\ref{eq:beta}). 

Whereas mass loss only weakly affects the GRB  luminosity, it
does affect the duration of the prompt emission: the larger the mass loss,
the earlier the MAD onset, the shorter the
GRB. Figures~\ref{fig:massloss1} and \ref{fig:massloss2} indicate the
MAD onset
time with blue stars: for a stronger mass-loss (larger 
values of $\beta$), the MAD forms at an earlier time.  This is to be expected,
since mass loss diminishes mass accretion rate and therefore BH
magnetic flux, which remains unaffected by mass loss, becomes
dynamically important at an earlier time.

Thus, all our previous conclusions hold in the presence of mass-loss. Summing up, jet power throughout
the GRB duration is rather constant. The presence of mass-loss causes
the remnant BH to be smaller, and
therefore, for the same available magnetic flux the jet power to be
slightly larger. Since mass-loss reduces $\dot M$ at the BH, the disk
turns MAD earlier and therefore the GRB duration is shortened.  The
effects of the mass-loss on GRB lightcurves are thus moderate at most.

\subsection{Effect of Stellar Rotation}
\label{sec:effect-stell-rotat}

An important, but poorly understood, property of GRB progenitor stars
is their rotation.  There is observational indication of
anti-correlation between stellar metallicity and the occurrence of
GRBs \citep[see, e.g.,][]{2008AJ....135.1136M}.  This can be
understood if rapid rotation of progenitor is required for its core
collapse to result in a GRB: higher metallicity leads to stronger
stellar winds, which extract stellar angular momentum and slow down
stellar rotation. In fact, there are many competing models, and the
underlying physical processes are not agreed upon.

Motivated by this, we explore the effect of stellar rotation on the
emergent GRB luminosity within our model.  We started by varying the
stellar rotation rate, $\omega$, by a constant factor relative to
$\omega_*(r)$, the rotation rate in our fiducial model. 
Figure~\ref{fig:pjetslowrot}(a) shows the analog of
Fig.~\ref{fig:pjet}(a) computed for $\omega = 0.23 \omega_*$. The
decrease in stellar rotation causes the jet to break out at a much
later time.  This is because the lower angular momentum of the stellar
envelope means that most of the star collapses into a BH directly,
bypassing the formation of an accretion disk. Since we assume that the
presence of an accretion disk is required for the production of jets
(see Sec.~\ref{sec:temp-evol-grb}), the formation of the disk and jets
does not take place until much later and can lead to a much shorter
duration of the GRB. For instance, in our fiducial model, disk (and
jet) formation takes place $\sim 10$~s after the core collapse (see
Fig.~\ref{fig:pjet}a), and the duration of the GRB is comparable to
the stellar free-fall time, or $\simeq 100$~s. However, for $\omega =
0.23\omega_*$, the disk forms at a much later time, $\approx 56$~s after
the core collapse and jet breaks out at $\approx 61$~s (somewhat later
for weaker $\Phi_*$, as seen in Fig.~\ref{fig:pjet}).  The duration of
the GRB is reduced, down to $57$~s. Thus, the slower the stellar
rotation, the shorter is the GRB.
For very slow stellar rotation, this effect would
prevent the standard GRBs from happening altogether: the GRB starts in
the steep decay stage and might instead appear as a low-luminosity GRB, as we
discuss below.

The above example makes it clear that the standard GRB lightcurve
profile is extremely resilient, even when the stellar rotation is
extremely slow: the jet power follows a plateau, which abruptly ends
in a steep decline. It could be a concern that the presence of
large-scale magnetic field in a progenitor star might lead to the
slow-down of the stellar core by magnetic torques and even
potentially result in solid-body--like rotation of the star as a
whole. Can GRBs be produced in such an unfavourable scenario? 

To explore this, we consider
solid-body rotation in the progenitor star, with a constant angular
velocity, $\omega = {\rm constant}$. We choose the rotation rate such
that the surface layers of the star rotate at $10$\% of Keplerian
value, which corresponds to surface rotational velocity $v_* \approx
200$~km~s$^{-1}$. 
As Fig.~\ref{fig:mdotsolidbodyrot}(a) illustrates,
the mass accretion rate is unaffected by the rotation profile 
(c.f.~Fig.~\ref{fig:mdot}a). Since for solid-body rotation most of the angular momentum is carried by the
outer layers of the star, the collapse of the core leads to 
essentially a non-spinning BH, $a(t = 1\ {\rm s})\approx 0.003\ll1$ (see
Fig.~\ref{fig:mdotsolidbodyrot}b), which is too low a spin to power a jet of
substantial power. Thus, naively, it would seem that this scenario is
hopeless for producing a GRB!

However, this is not so: just as we saw above in the case of slow
rotation, the GRB does not start until the infalling gas hits the
centrifugal barrier and forms an accretion disk and a jet. In this
model, this happens around $t\approx 57$~s (marked with red circle in
Fig.~\ref{fig:mdotsolidbodyrot}a), and the jet breaks out of the star
and the GRB starts at $t\approx 76$~s (indicated
by the green square). By this late time, the central BH receives most
of the angular momentum carried by the outer layers of progenitor
star, and BH spin saturates at a respectable value, $a\approx0.4$. Jet
power, shown in Fig.~\ref{fig:pjetsolidbodyrot}(a), and BH magnetic
flux, shown in Fig.~\ref{fig:pjetsolidbodyrot}(b), saturate at
near-constant values standard for GRBs: $L_j \lesssim
10^{50}$~erg~s$^{-1}$ and $\Phi\sim10^{27}\ \Gcmsq$. The end of the
GRB is marked by the onset of MAD and occurs at approximately the same
time as in our fiducial model, $t_{\rm MAD}\approx 100$~s. Thus, the
duration of the GRB is shortened down to $\simeq40$~s from
$\simeq100$~s in our fiducial model. 

Thus, the details of radial distribution of
angular momentum inside the star are not important for the
observational appearance of the GRB: for instance, GRB light curves
are essentially identical for the slow stellar rotation, with $\omega
= 0.23\omega_*$ (see Fig.~\ref{fig:mdotslowrot}) and the case of
solid-body rotation, with $v_* = 200$~km~s$^{-1}$ (see
Fig.~\ref{fig:mdotsolidbodyrot}). The only major underlying
difference is that the latter results in an approximately twice as
low BH spin and therefore a somewhat shorter GRB. For instance, for
the weakest stellar magnetic flux value we considered, $\Phi_*=
10^{26.5}$~G~cm$^2$, the jet does not break out of the star at all
(see Fig.~\ref{fig:mdotsolidbodyrot}a,b) and thus no normal GRB is
produced. Therefore, for a normal GRB to occur, it is important
to have both sufficient stellar magnetic flux and stellar angular momentum.

The slower the rotation, the shorter the GRB duration. As the angular
momentum of the progenitor is reduced, the duration of a core-collapse
GRB can become shorter than $\simeq2$~s, which is the standard divide
between the populations of short- and long-duration GRBs
\citep{kouv93}. This can make long-duration GRBs appear as short GRBs
\citep[see also][]{2012ApJ...749..110B,2013ApJ...764..179B}.  For very
slow rotation, the jets either barely have enough time to break out of
the star 
or fail to break
out of the star and can result in a low-luminosity
GRB or an X-ray flash. If the analogy with low-luminosity GRBs holds,
such a failed or near-failed GRB can be accompanied by a supernova
explosion \citep{stanek03,2006ApJ...645L..21M}.

So far we have assumed that the for the jet to form, the accreting
material needs to hit a centrifugal barrier.  What happens if the gas
never hits the centrifugal barrier and the disk does not form?  In
such a limit of very slow rotation, it might seem that the jet never
forms and there would be no GRB at all, as without the disk, there
is no low-density funnel through which the jet can
escape. However, this is not the case. In fact, the low-density funnel
is only a necessity for jet formation when the magnetic field is
\emph{not} dynamically-important. However, once a MAD forms, a jet will be
produced even by a non-rotating accretion flow in the absence of a
low-density funnel
\citep{kb09,2014MNRAS.437.2744T}. We estimate that the energetics of
such events is of order $L_{\rm MAD} t_{\rm MAD}/20$, or about $5\%$
of the energetics of normal GRBs. (Here the factor of $1/20$ comes from
integrating the luminosity time-dependence $L_j \propto t^{-20}$
around the onset of the
MAD regime.)  Thus, on the energetics grounds it is conceivable that such events could indeed be
the counterparts of low-luminosity GRBs.

Summarising, GRBs appear to be successfully produced over a wide range
of progenitor rotation rates and profiles. The lower the angular
momentum of the progenitor star, the shorter is the GRB that results.
However, regardless of the progenitor rotation rate and profile, GRBs
robustly show an early-time plateau followed by the sharp
decline. Thus, the observational appearance of GRBs appears to be
insensitive to the details of angular momentum distribution inside the
progenitor star. How do other parameters of the progenitor affect the
GRB?

\begin{figure}
\begin{center}
    \includegraphics[width=0.72\columnwidth]{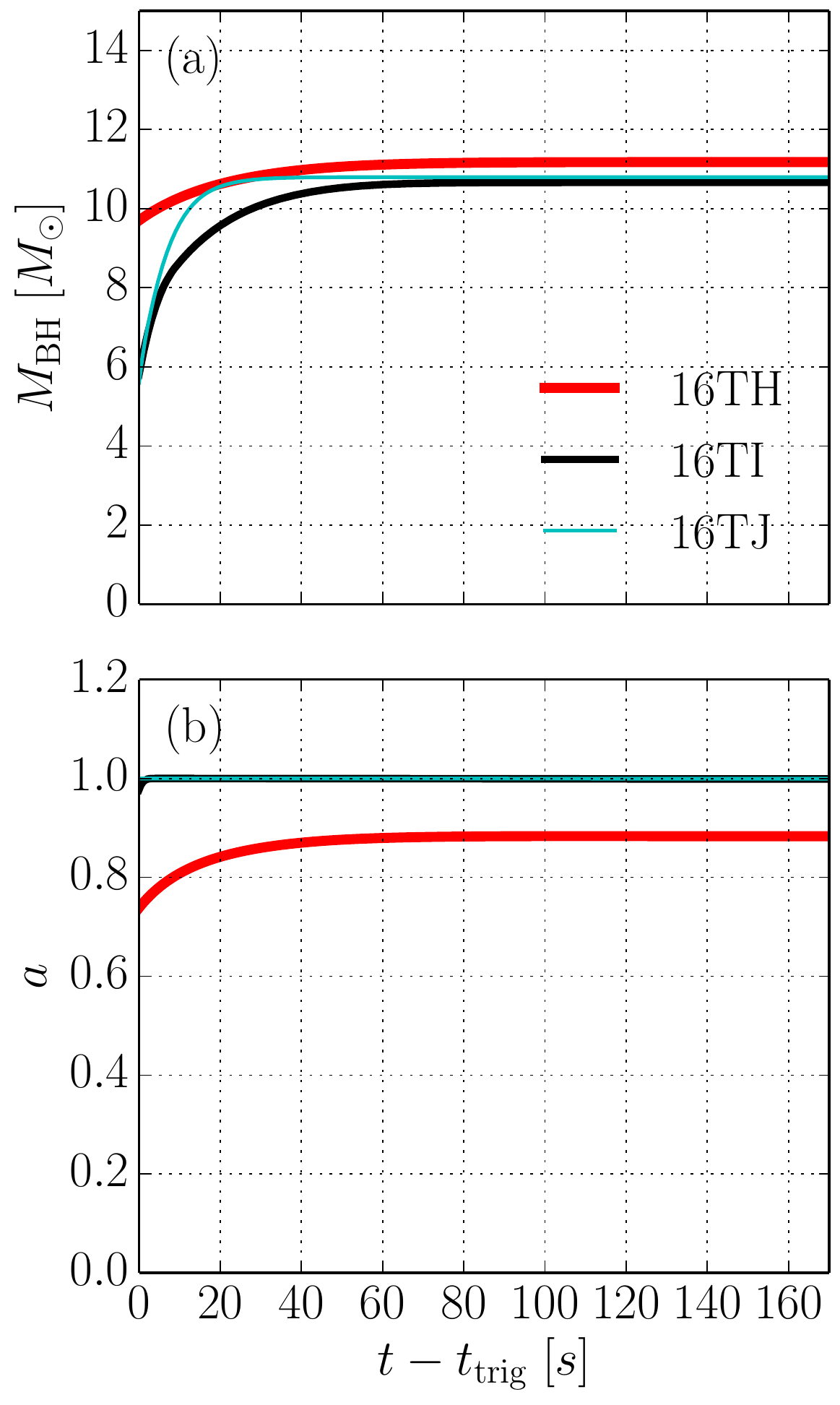}
\end{center}
    \caption{BH mass and spin vs. time since the trigger for different
      stellar progenitor models: 16TH (thick solid red lines), 16TI
      (medium-thickness solid black lines), and 16TJ (thin cyan
      lines).  In all cases, we use our
  fiducial model parameters: $\gamma=1$, $\beta=0$,
  $\Phi_*=10^{27.5}$. Panel (a) shows that BH mass change is between $10\%$
      for model 16TH and $50\%$ for models 16TI and 16TJ. BH spin remains approximately constant in time (to
      better than 20\%) and levels off to a high value: $\simeq 0.9$
      for model 16TH and $\simeq1$ for models 16TI and 16TJ.}
\label{fig:Mamodels}
\end{figure}

\begin{figure}
\begin{center}
    \includegraphics[width=0.9\columnwidth]{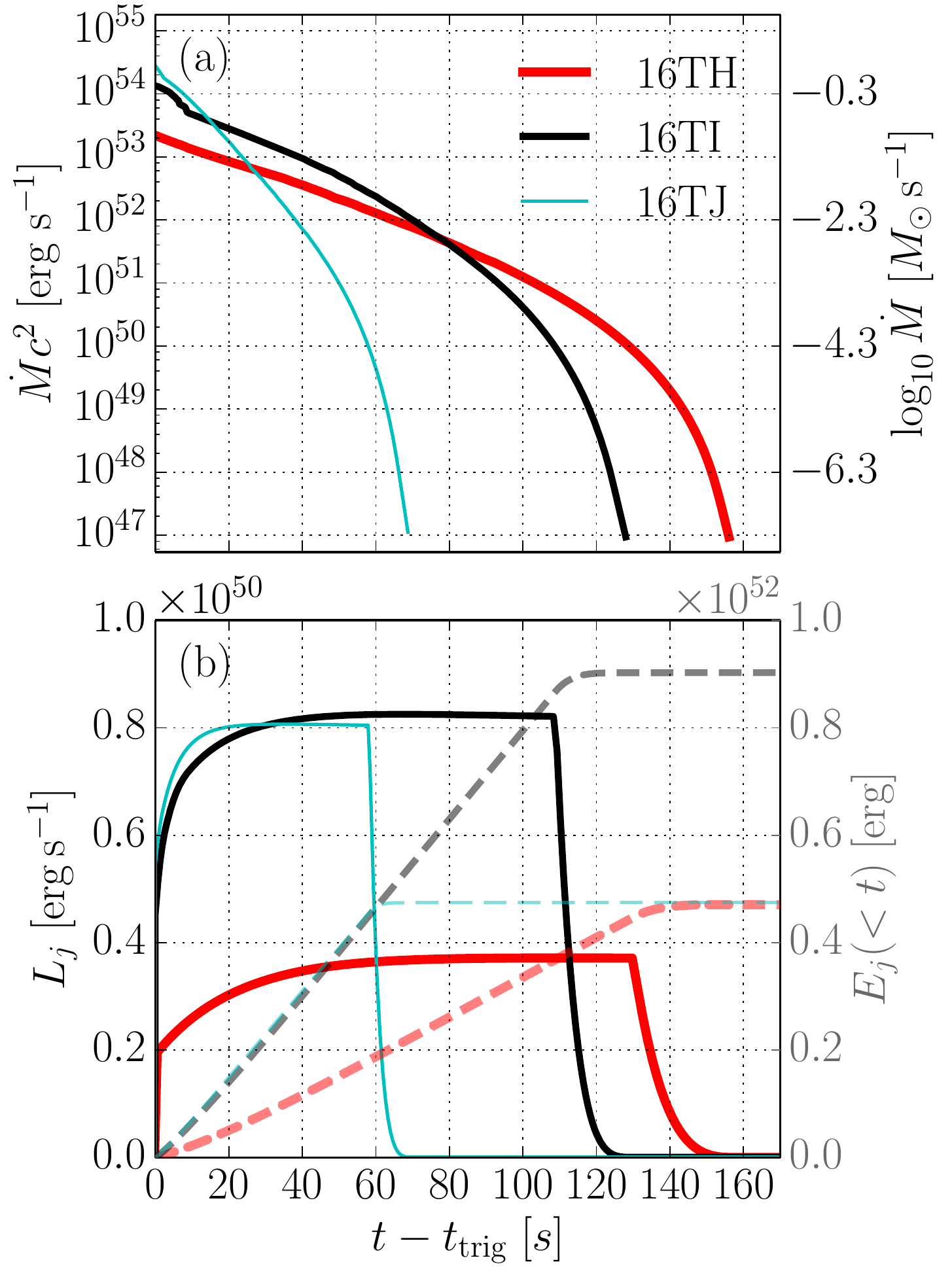}
\end{center}
\caption{Time-variation of various quantities for different stellar
  progenitor models (see the legend and caption to
  Fig.~\ref{fig:Mamodels} for details). In all cases, we use our
  fiducial model parameters: $\gamma=1$, $\beta=0$,
  $\Phi_*=10^{27.5}$. {\bf [Panel (a)]:} shows mass accretion rate
  $\dot M$ on a \emph{logarithmic scale}: it changes by many orders of
  magnitude during the GRB and differs from one model to another by
  $1-2$ orders of magnitude. {\bf [Panel (b)]:} Solid lines show 
  the power of the
  jets $L_j$ on a \emph{linear scale}. In contrast to $\Mdot$, the power remains
  relatively constant and changes between the models by at most a
  factor of $2$. This is because during the GRB, jet power is
  decoupled from the widely varying mass accretion rate. To help us
  visually quantify the constancy of $L_j$, the lighter-coloured dashed
  lines and the right $y-$axis show cumulative light curves of jet
  power, $E_j(<t)=\int_0^tL_j dt'$, for the three models. That $E_j(<t)$
  shows little curvature throughout the prompt emission is a
  reflection of the near-constancy of its slope, or jet power. }
\label{fig:MPmodels}
\end{figure}

\subsection{Effect of Stellar Progenitor Model}
\label{sec:effect-prec-model}

We will now investigate the effect of the pre-collapse stellar model on GRB
light curves. For simplicity, in all cases we choose our fiducial
parameters: $\gamma = 1$, $\beta = 0$ ($=$no mass loss), and
$\Phi_* = 10^{27.5}$. Figure~\ref{fig:Mamodels} shows the 
time-dependence of BH mass (panel a) and spin (panel b) for 3
different progenitor models: 16TH, 16TI and 16TJ. 
The temporal profiles of $\MBH$
and $a$ are qualitatively similar in all 3 models: both mass and
spin level off early in the course of the GRB and do not change
thereafter. In all models BH spin ends up at a rather high value,
$a\gtrsim0.9$.

Figure~\ref{fig:MPmodels}(a) shows that $\Mdot$ drops by several
orders of magnitude during the GRB.  However, $\Mdot$ does not
directly control jet power, $L_j$ (see
Sec.~\ref{sec:bhmadestimate}). For this reason, $L_j$, remains mostly constant
throughout the GRB, as is clear from Fig.~\ref{fig:MPmodels}(b).
All 3 progenitor models lead to virtually indistinguishable
shapes of prompt emission light curves: all of them show about $50\%$
variation at the very beginning of the GRB, after which the emission
levels off to a constant. In all cases, at the end of the GRB, the jet
power abruptly drops. This happens when $\Mdot$ drops below
$\Mdot_{\rm MAD}$ (see eq.~\ref{eq:mdotmad}), after which jet power
begins to track the rapidly decreasing mass accretion rate.  The
durations of the GRBs for the 3 different stellar models 
we have considered are within a factor of $2$ of each other and in
all cases are set by the free-fall time scale of the outer layers of
the progenitor star.

In summary, all basic properties like $\MBH$, $a$, and $\Mdot$ do not
differ substantially between different stellar models.  For a fixed magnetic
flux through the progenitor, $\Phi_*$, the jet power
$L_j$ differs by factors of just a few.  The magnetic
flux therefore appears to be the only quantity that can feasibly
account for the huge range of observed GRB luminosity. Note, however,
that the progenitor properties such as rotation can have an effect on
the GRB duration (Sec.~\ref{sec:effect-stell-rotat}) and may have an
indirect effect on the field strength (e.g.\ the faster rotators can
have stronger magnetic fields).  Since the origin of stellar magnetism
is not well-understood, we do not explore such cross-correlations in
this work.

\section{Comparison with GRB lightcurves}
\label{sec:comparison-with-grb}

The GRB emission is made up of a superposition of many
pulses.  The pulse properties, such as peak flux, duration,
inter-pulse intervals, do not change in any systematic way during the
burst. As a result, cumulative photon counts during GRBs increase, on
average, linearly with time: $\int_0^t L_\gamma{\rm d}t\sim {\rm
  constant}\times t$ or
$L_\gamma\sim$ constant \citep{2002A&A...393L..29M}. Here we show that
our model reproduces this observation.

\begin{figure}
\begin{center}
    \includegraphics[width=\columnwidth]{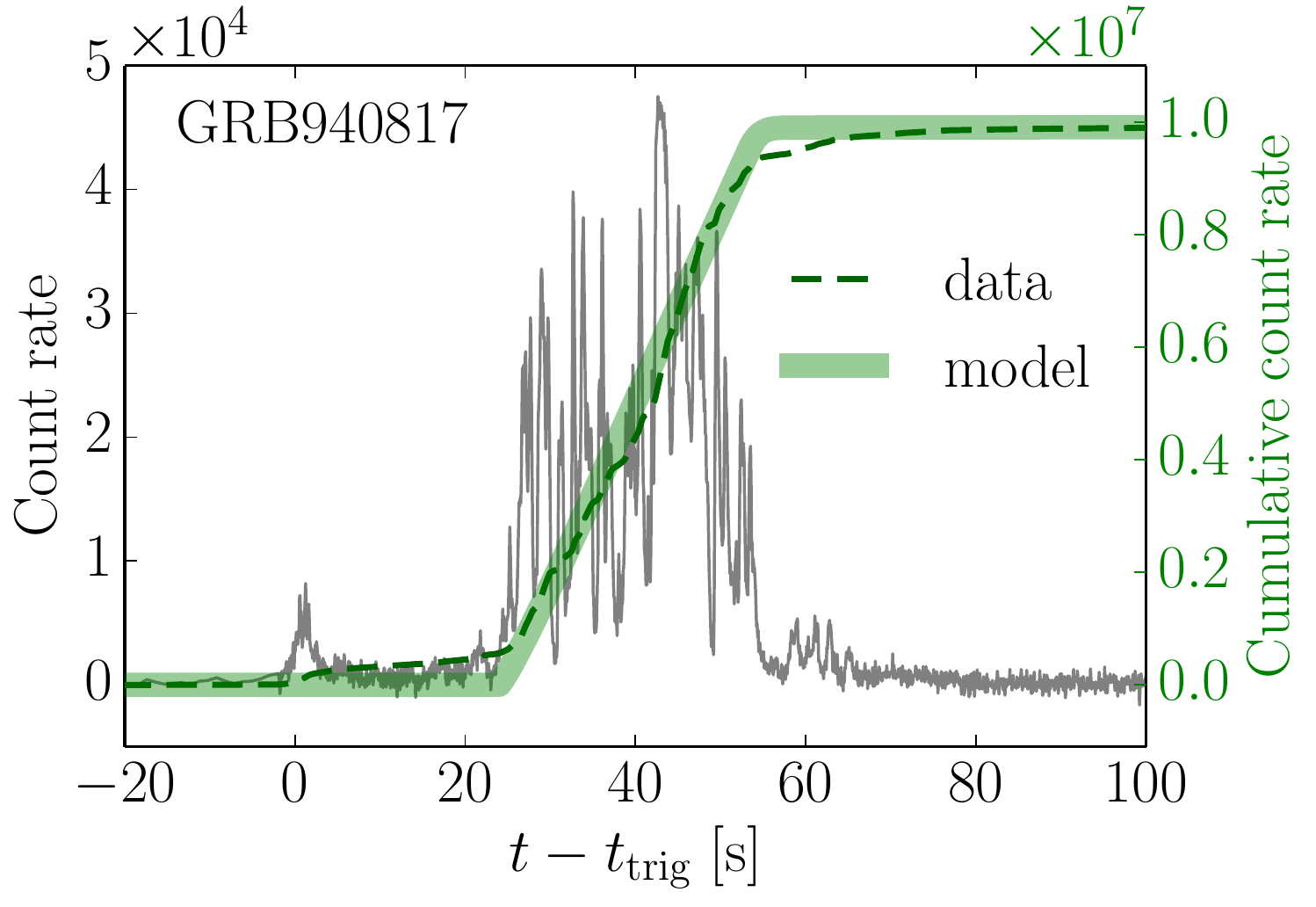}
\end{center}
  \caption{BATSE X-ray lightcurve (in all channels combined) for GRB
    940817 is shown
    with solid grey line (left $y-$axis). The cumulative light curve
    is shown with dashed green line (right $y-$axis). Light green stripe shows the model-predicted cumulative lightcurve for progenitor 16TI
    \citep{woosley_progenitor_2006} and the magnetic flux distribution
    index of $\gamma = 1$.}
\label{fig:lc3128}
\end{figure}

\begin{figure}
\begin{center}
    \includegraphics[width=\columnwidth]{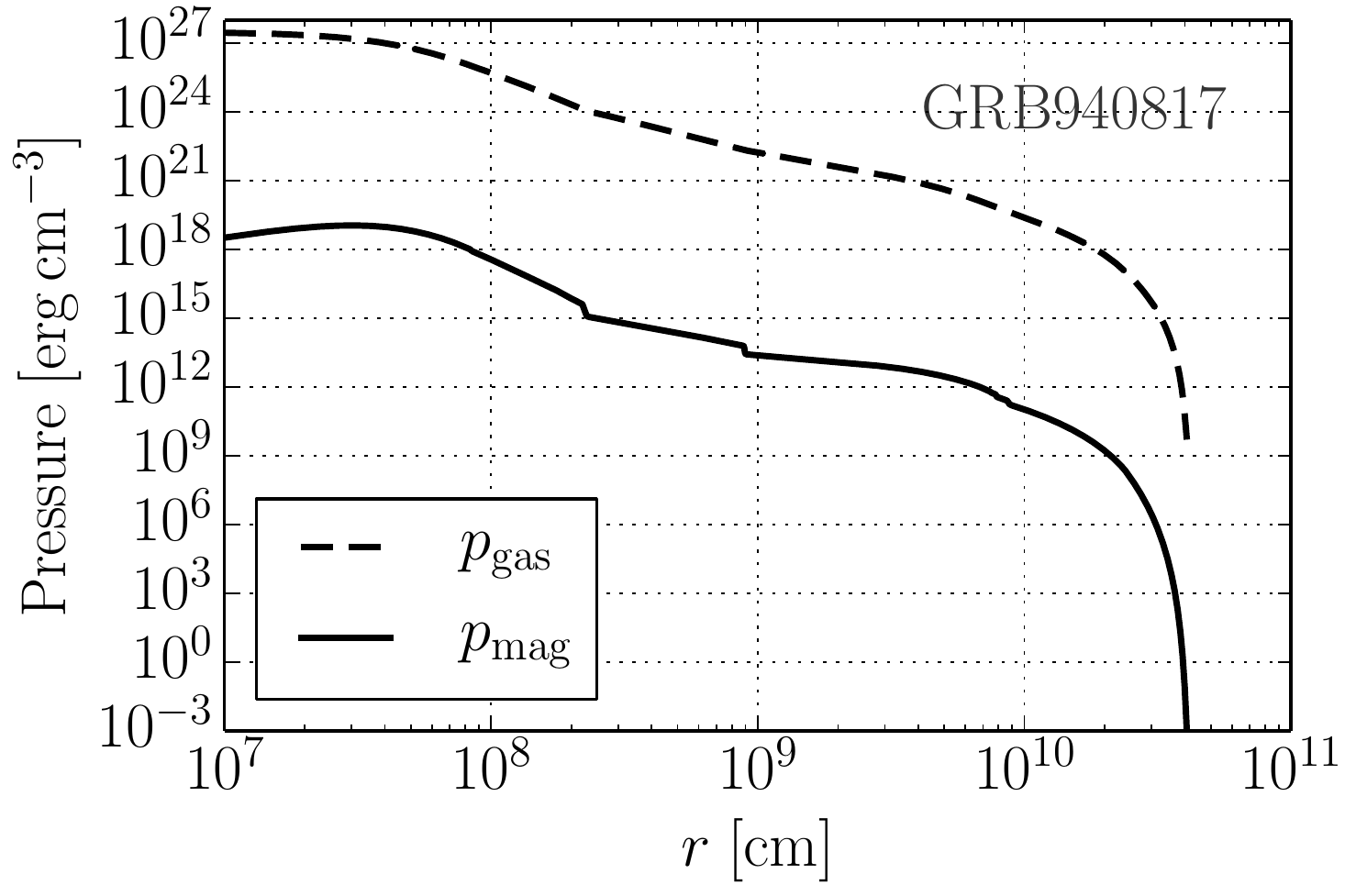}
\end{center}
  \caption{The radial profile of gas and magnetic pressures
    corresponding to Fig.~\ref{fig:lc3128}.}
\label{fig:lc3128p}
\end{figure}

In Fig.~\ref{fig:MPmodels}(b) we plot along the right
$y-$axis the cumulative light curves of our simulated GRBs with dashed
lines. The cumulative light curves are mostly straight lines
throughout the prompt emission, indicating the near-constancy of jet
power. We will now apply this type of visual analysis to several
representative BATSE light curves.

\citet{2002A&A...393L..29M} demonstrated that the cumulative count
rate increases linearly with time in a sample of $\sim 500$ bright
BATSE bursts.  Here we repeat their analysis for representative GRBs,
940817, 940210, 920513, and compare the results to the predictions
of our model.

\begin{figure}
\begin{center}
    \includegraphics[width=\columnwidth]{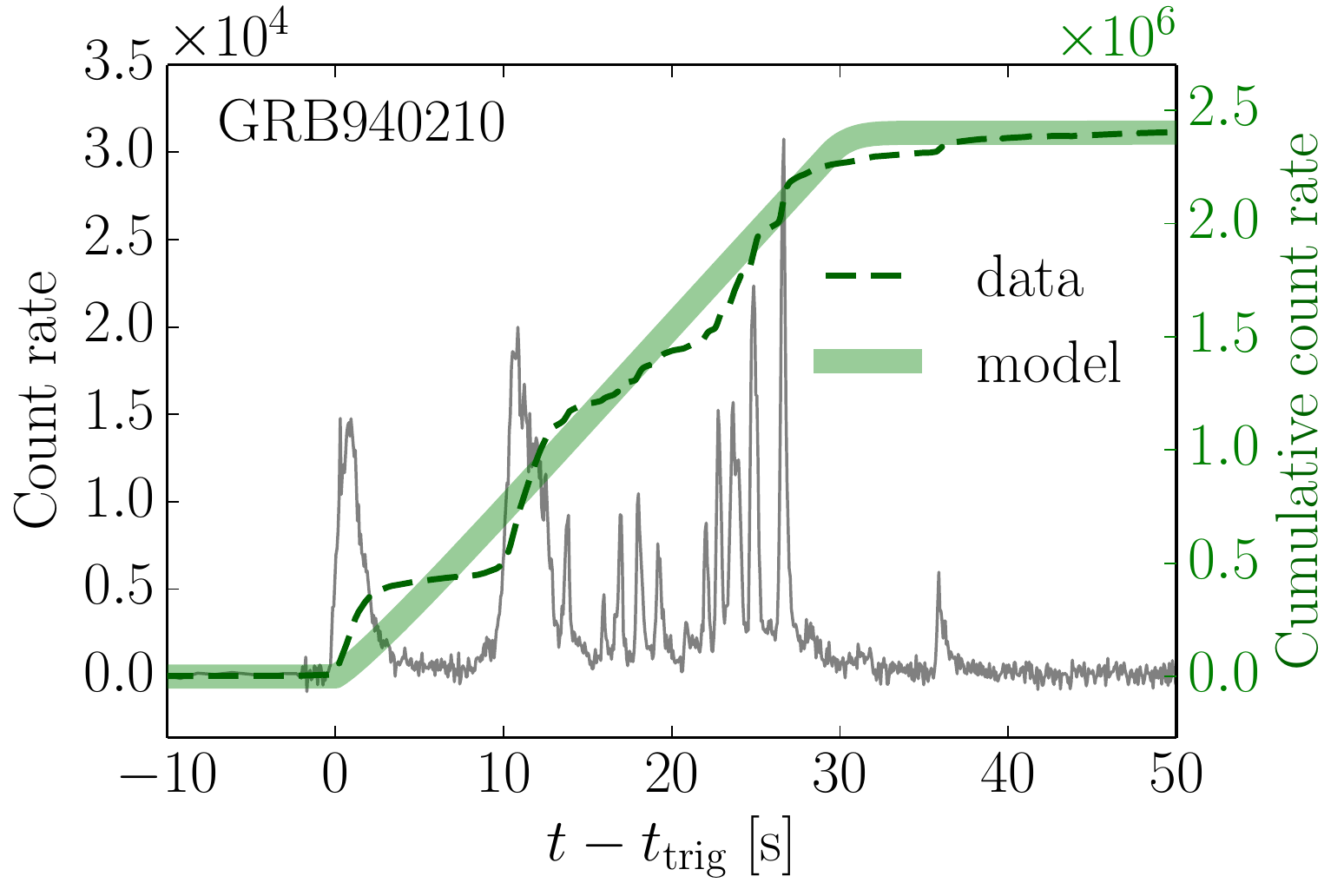}
\end{center}
  \caption{BATSE X-ray lightcurve (in all channels combined) for GRB
940210 is shown
    with solid grey line (left $y-$axis). The cumulative light curve
    is shown with dashed green line (right $y-$axis). Light green stripe shows the model-predicted cumulative lightcurve for progenitor 16TI
    \citep{woosley_progenitor_2006} and the magnetic flux distribution
    index of $\gamma = 1$.}
\label{fig:lc2812}
\end{figure}

We chose GRBs 940817 and 940210 because they are rather clean examples of the
linear increase of the radiated energy versus time followed by a sharp
turn-off of the GRB emission. In Figs.~\ref{fig:lc3128}
and~\ref{fig:lc2812} we overplot the cumulative count rate from these
bursts with those for our model (we rescaled the time axis to fit the
observed duration). We adopt our fiducial model parameters: the
magnetic flux distribution index of
$\gamma=1$, the total magnetic flux of $\Phi_* = 10^{27.5}$~G~cm$^2$, no
mass-loss from the disk ($\beta = 0$), along with the
16TI progenitor star model. The
results for other combinations of model parameters, $\gamma$, $\beta$,
and $\Phi_*$, and progenitor models are similar (not shown).

\begin{figure}
\begin{center}
    \includegraphics[width=\columnwidth]{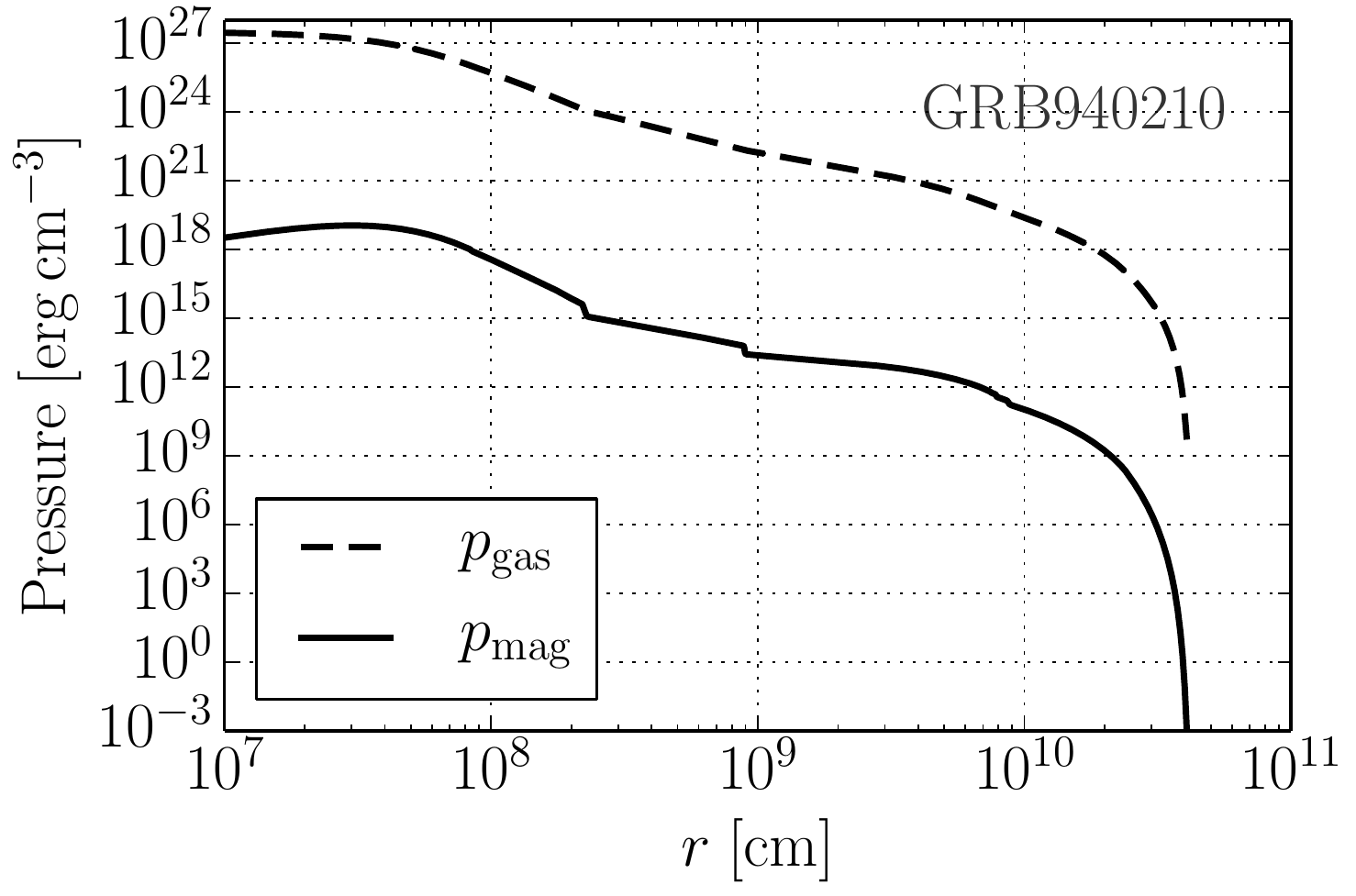}
\end{center}
  \caption{The radial profile of gas and magnetic pressures
    corresponding to Fig.~\ref{fig:lc2812}.}
\label{fig:lc2812p}
\end{figure}

As is illustrated by Figs.~\ref{fig:lc3128} and \ref{fig:lc2812}, our
models naturally account for the near constancy of the burst
luminosity over its duration as well the steep decline of the jet
power at $t-t_{\rm trig}\sim 60$~s and 30~s, respectively. For
comparison, in Figs.~\ref{fig:lc3128p} and \ref{fig:lc2812p} we show
the implied gas and magnetic pressures as functions of distance in the
pre-collapse star for the same two models. The gas pressure dominates
magnetic pressure in the star by a typical factor of $\sim 10^9$.


\subsection{Magnetic tomography of progenitor stars}
\label{sec:magn-tomogr-prog}

\begin{figure}
\begin{center}
    \includegraphics[width=\columnwidth]{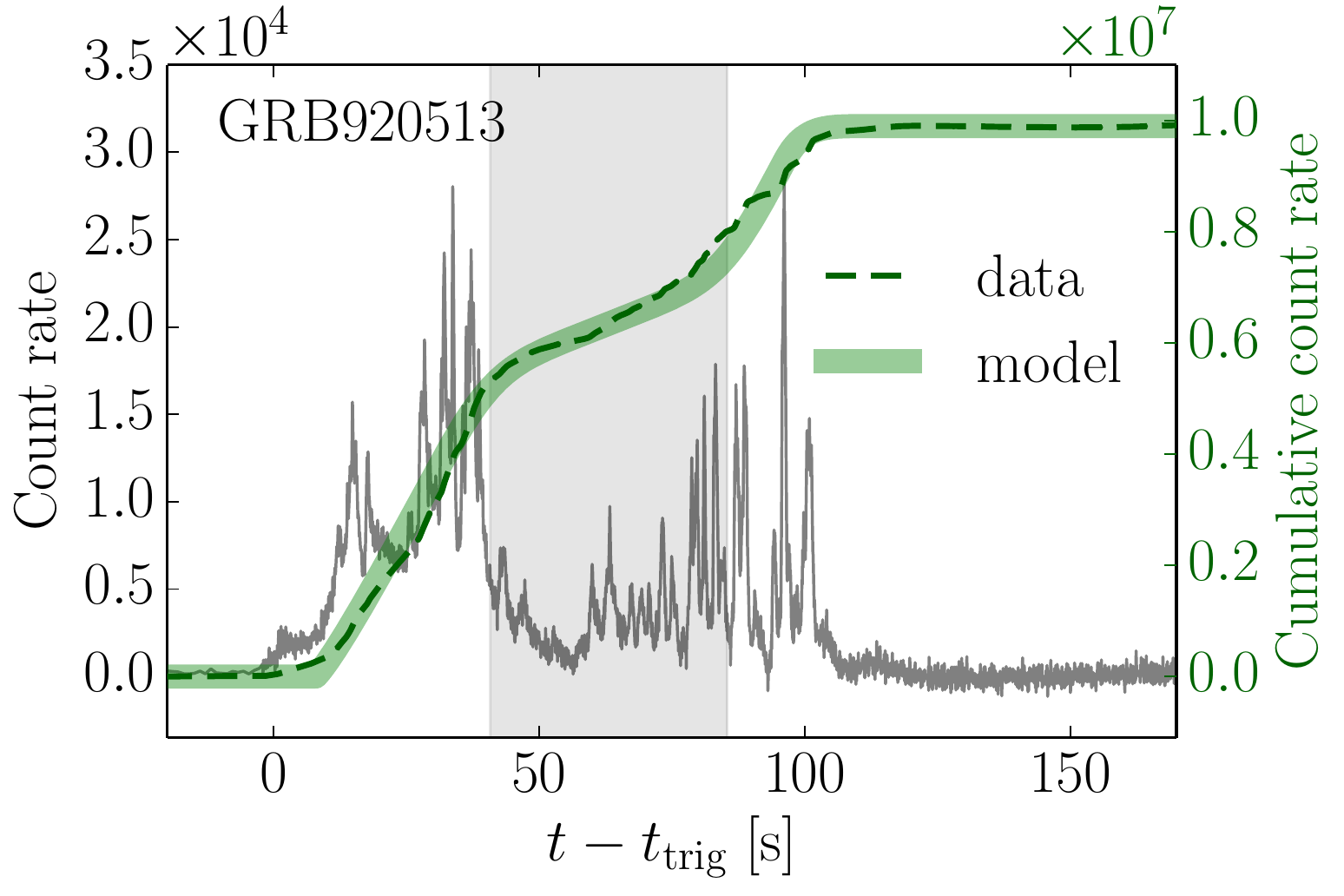}
\end{center}
  \caption{BATSE X-ray lightcurve (in all channels combined) for GRB
920513 is shown
    with solid grey line (left $y-$axis). The cumulative light curve
    is shown with dashed green line (right $y-$axis). Light green stripe shows the model-predicted cumulative lightcurve for progenitor 16TI
    \citep{woosley_progenitor_2006} and the magnetic flux distribution
    index of $\gamma = 1$. To model the suppression of power in the
    second half of the GRB, we added a torus-like magnetic field
    component centred at $t-t_{\rm trig}= 70$~s and indicated with
    the vertical grey stripe. The magnetic field in the inner part of
    the torus is of opposite
    polarity to that in the rest of the star. Hence, the accretion of
    the torus causes a temporary suppression of GRB power.
}
\label{fig:lc1606}
\end{figure}

\begin{figure}
\begin{center}
    \includegraphics[width=\columnwidth]{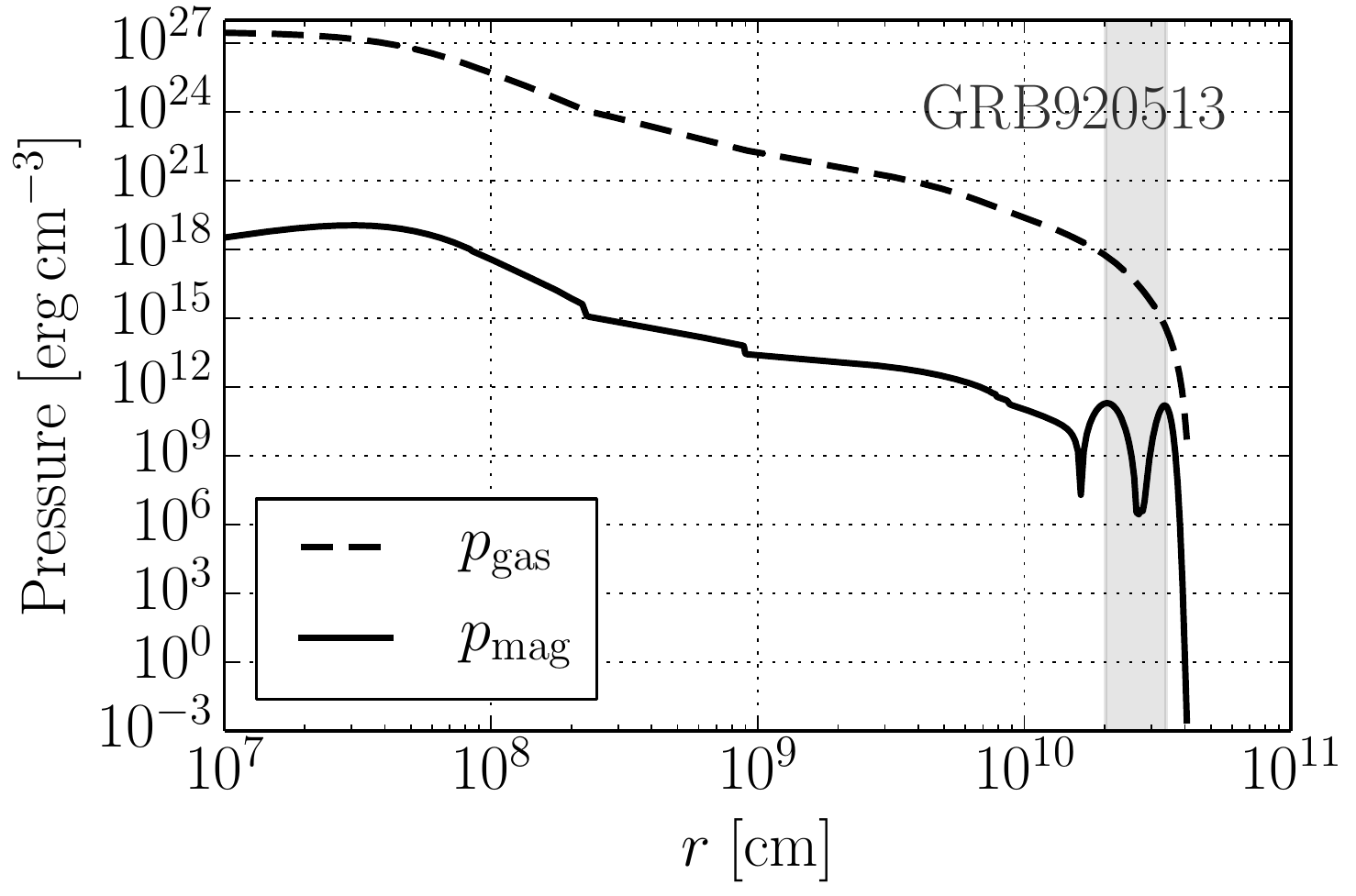}
\end{center}
\caption{The radial profile of gas and magnetic pressures
  corresponding to Fig.~\ref{fig:lc1606}. For this burst, we added an
  additional magnetic torus of opposite polarity to that in the rest
  of the star.  This component is centred at $t - t_{\rm trig}\sim 60$~s
  (approximately at $r\sim 3\times10^{10}$~cm), with the width of
  $\sim20$~s (approximately $10^{10}$~cm), and indicated with the
  vertical grey stripe.  It appears as two bumps in
    the magnetic pressure profile (see Fig.~\ref{fig:lc1606} and text for details).}
\label{fig:lc1606p}
\end{figure} 

From Fig.~\ref{fig:lc1606} it is clear that gross properties of the
cumulative count rate from GRB 920513 can also be reproduced by this
simplest version of the model.  However, the cumulative light curve
has either a pronounced bump at $t-t_{\rm trig}\sim 40$~s or a deficit
at $t-t_{\rm trig}\sim 60$~s.  

Such a deviation from the simple prescription where magnetic flux
scales with mass in the progenitor star can be used as a probe of the
stellar field structure. To account for the relative weakness of the
burst at the interval $50\lesssim t\lesssim 80$s, we introduce an
additional torus-shaped component of poloidal magnetic field centred
at $t-t_{\rm trig} \sim 60$~s, or a distance $r\sim 3\times 10^{10}$cm in
the progenitor star. This component appears in Fig.~\ref{fig:lc1606p} as a pair
of bumps in the magnetic pressure and is indicated by a vertical grey band.
Namely, we superimpose on top of our $\gamma = 1$
magnetic flux profile a magnetic ``knot'': a torus-like poloidal
magnetic field component of negative polarity, such that the
consumption by the BH of this knot temporarily depresses BH magnetic flux. We take the
time-dependence of BH magnetic flux of the following form: 
\begin{equation}
\PhiBH^{\rm GRB920513}(t) = \Phi_*\left(\frac{M_{\rm
  collapsed}}{M_*}\right)
\left\{1 -f_0\exp\left[-
    \left(\frac{t-t_{\rm trig}-t_0}{\Delta t_0}\right)^4
\right]\right\},
\label{eq:PhiBHtorus}
\end{equation}
where the term in the curly braces represents BH magnetic flux
suppression factor. For illustration, we performed a by-eye fit and
obtained the following parameters $f_0 = 0.55$, $t_0 = 63$~s, and
$\Delta t_0 = 23$~s.  Here, $f_0$ is a dimensionless
parameter equal to the dimensionless magnetic flux carried by the
knot, i.e., the ratio of the magnetic flux in the knot to the net flux
in the star. Thus, $f_0 = 1$ means that, as the knot is consumed by
the BH, the jet power vanishes
completely, a value of $f_0 \lesssim 1$ leads to a substantial
suppression of BH magnetic flux and GRB luminosity, whereas $f_0\ll 1$
leads to a small dip in the flux and luminosity.

When the inner half of the
magnetic knot accretes ($t-t_{\rm trig}\sim t_0-\Delta t_0 = 40$~s),
the total BH magnetic flux gets reduced by a factor of
$(1-f_0)^{-1}\approx2$ and the jet power by a factor of $(1-f_0)^{-2}
\approx 5$. Subsequent accretion of the outer half of the knot (at
$t-t_{\rm trig}\sim t_0+\Delta t_0 = 86$~s) replenishes the BH
magnetic flux and leads to the recovery of jet power. Note that the
magnetic knot fully resides under the stellar surface and so does not
affect the surface magnetic field strength.

\begin{figure}
\begin{center}
    \includegraphics[width=\columnwidth]{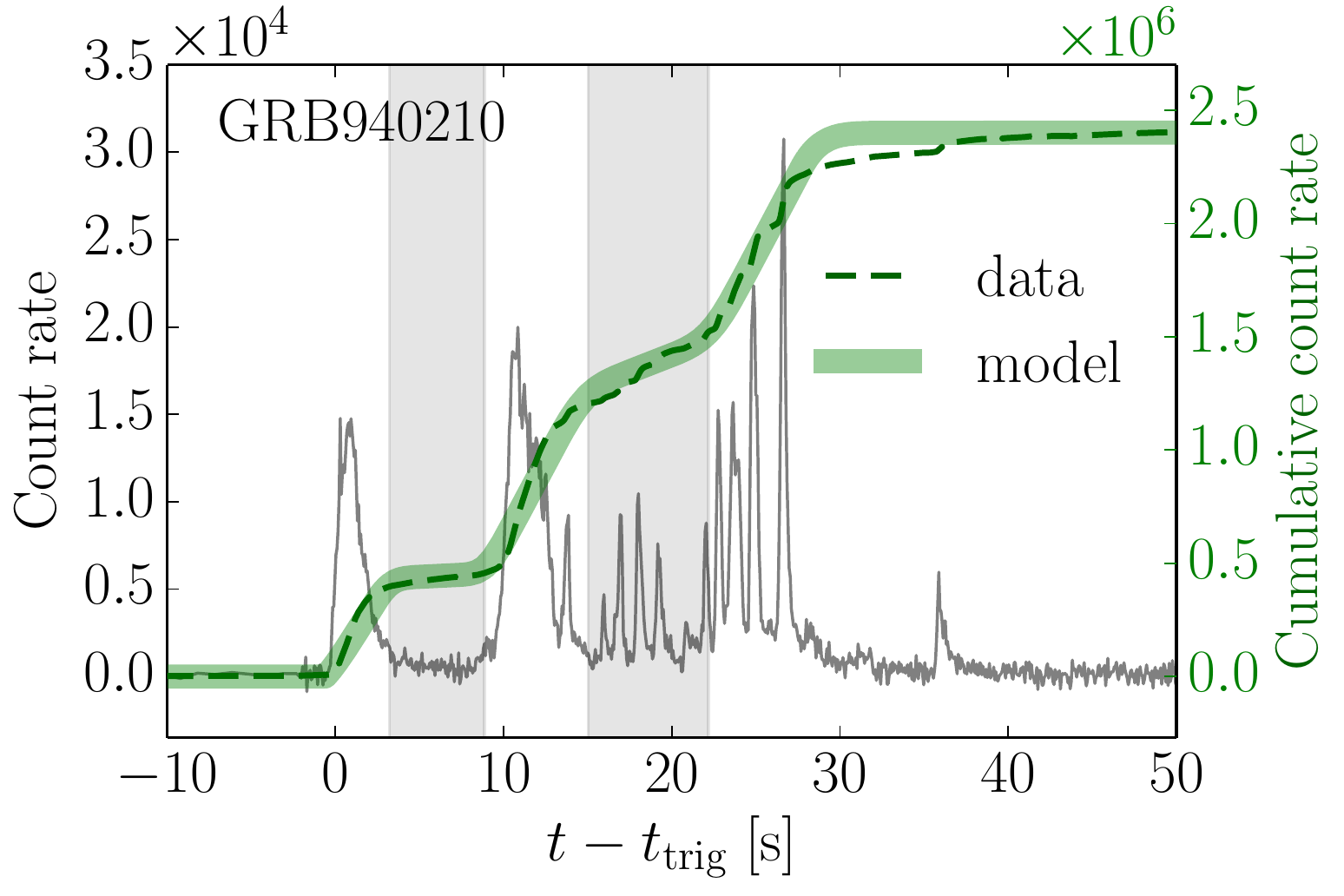}
\end{center}
\caption{Similar to Fig.~\ref{fig:lc2812} but with the addition of two
  magnetic knots, which are centred at $t-t_{\rm trig}\sim5$~s and $\sim
  20$~s and indicated by vertical grey stripes and containing magnetic
  fluxes that make $\sim85$\% and $\sim55$\% that through the star. Stellar magnetic flux
  distribution versus radius is given by eq.~(\ref{eq:PhiBHtorus2}). 
  As these knots are consumed by the BH, BH magnetic
  flux and GRB power are depressed. }
\label{fig:lc2812knots}
\end{figure}

\begin{figure}
\begin{center}
    \includegraphics[width=\columnwidth]{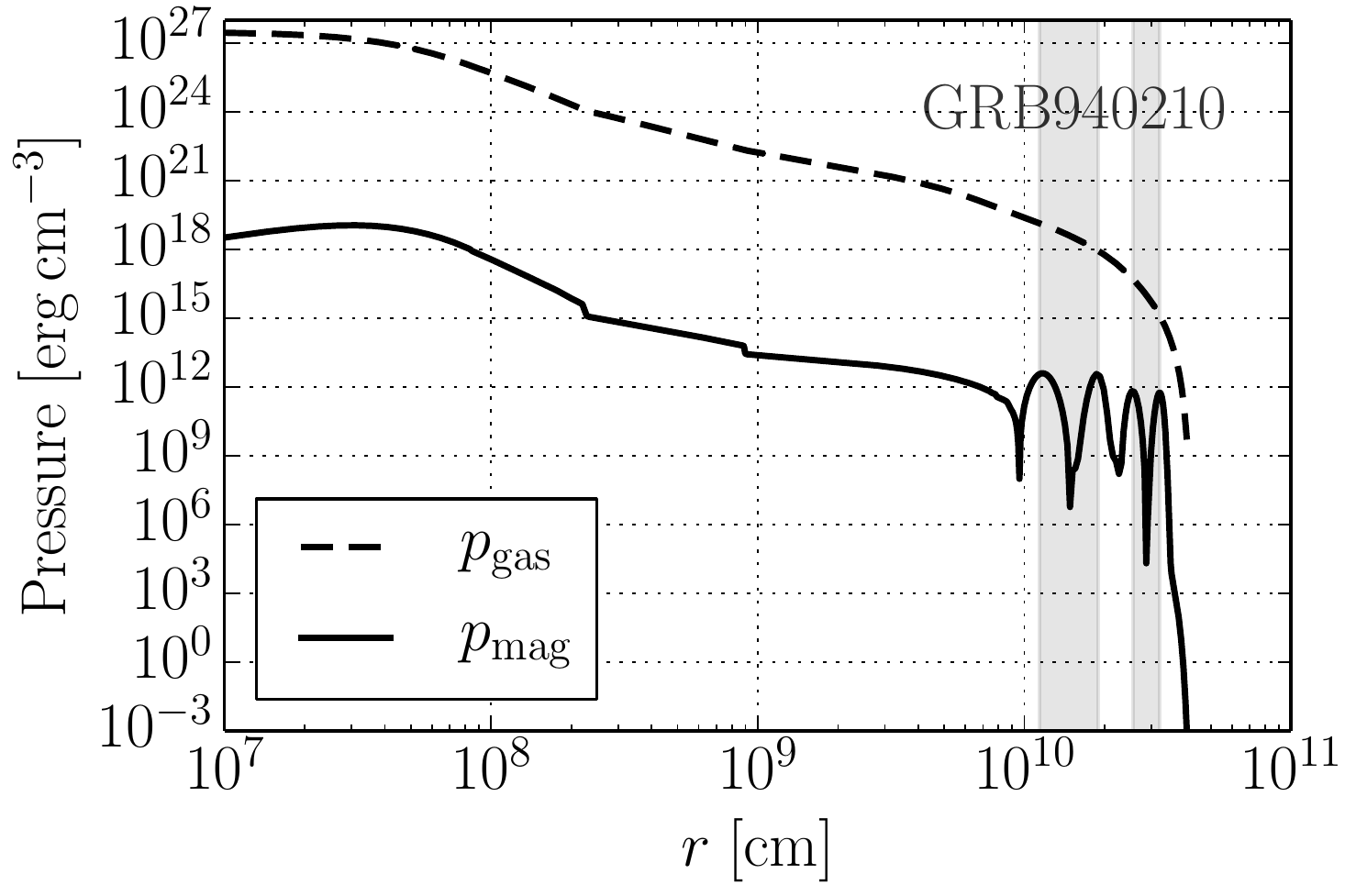}
\end{center}
  \caption{The radial profile of gas and magnetic pressures
    corresponding to Fig.~\ref{fig:lc2812knots}. Each of the two magnetic
    knots is seen as a pair of bumps on top of the
    underlying, monotonically decreasing magnetic pressure profile
    (which is shown in Fig.~\ref{fig:lc2812p}). Vertical grey stripes
  indicate the one-sigma extent of the knots that matches the extent of
  the bumps in magnetic pressure.}
\label{fig:lc2812knotsp}
\end{figure}

GRB 940210 lightcurve shows two prominent deficits, or quiescent
intervals, as seen in Fig.~\ref{fig:lc2812}: one at $t-t_{\rm trig}\sim5{-}10$~s and another one at
$t-t_{\rm trig}\sim15{-}20$~s. Similar to
GRB 920513, we can model these quiescent intervals by introducing
magnetic knots into an otherwise large-scale magnetic flux distribution
in the star. As each magnetic knot is accreted by the BH, the knot's
magnetic flux cancels out a substantial fraction of BH magnetic flux
and leads to a depression in GRB luminosity.  We model the
magnetic flux in the knots similar to eq.~\eqref{eq:PhiBHtorus}:
\begin{equation}
\PhiBH^{\rm GRB940210}(t) = \Phi_*\left(\frac{M_{\rm
  collapsed}}{M_*}\right)\times
\prod_{i=1,2}\left\{1 -f_i\exp\left[-
    \left(\frac{t-t_{\rm trig}-t_i}{\Delta t_i}\right)^4
 \right]\right\}.
\label{eq:PhiBHtorus2}
\end{equation}
Here we choose $f_1 = 0.85$, $t_1 = 6.1$~s, $\Delta t = 2.9$~s, and
$f_2 = 0.55$, $t_2 = 18.6$~s, $\Delta t_2 = 3.9$~s (we give the
precise values for reproducibility; however, qualitative
behaviour of the lightcurve is insensitive to the details of the fit). 
Thick green line in Figure~\ref{fig:lc2812knots} shows the
predicted cumulative GRB lightcurve with the thick solid green line:
the accretion of magnetic knots, which are indicated by grey vertical
stripes in Figs.~\ref{fig:lc2812knots} and~\ref{fig:lc2812knotsp},
leads to the partial depressions in BH magnetic flux and in turn to
quiescent intervals in jet power, similar to those seen in GRB
lightcurves.


\begin{figure}
\begin{center}
    \includegraphics[width=\columnwidth]{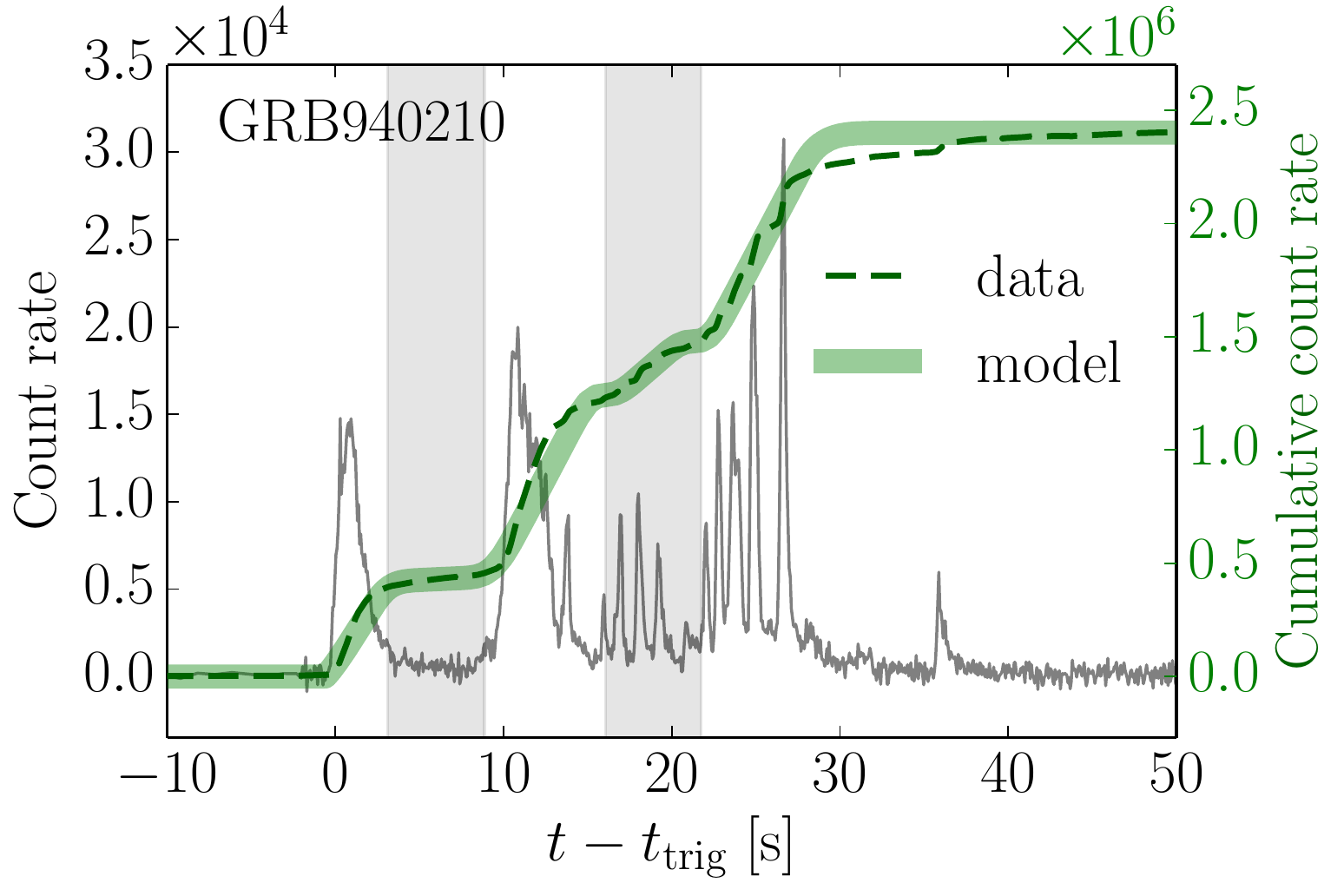}
\end{center}
\caption{Similar to Fig.~\ref{fig:lc2812knots} but with the two knots
  containing magnetic fluxes that make up $85$\% and $160$\% of those
  through the star (see text for details). Note the qualitative
  difference from Fig.~\ref{fig:lc2812knots} in which the knots
  contain $55$\% and $85$\% of that through the star: here we have a
  flux of $160\%>100$\%, which leads to ``flat--steep--flat'' behaviour of the
  cumulative light curve. This behaviour appears to be well-captured by our
  model (see text for details). }
\label{fig:lc2812knots2}
\end{figure}
\begin{figure}
\begin{center}
    \includegraphics[width=\columnwidth]{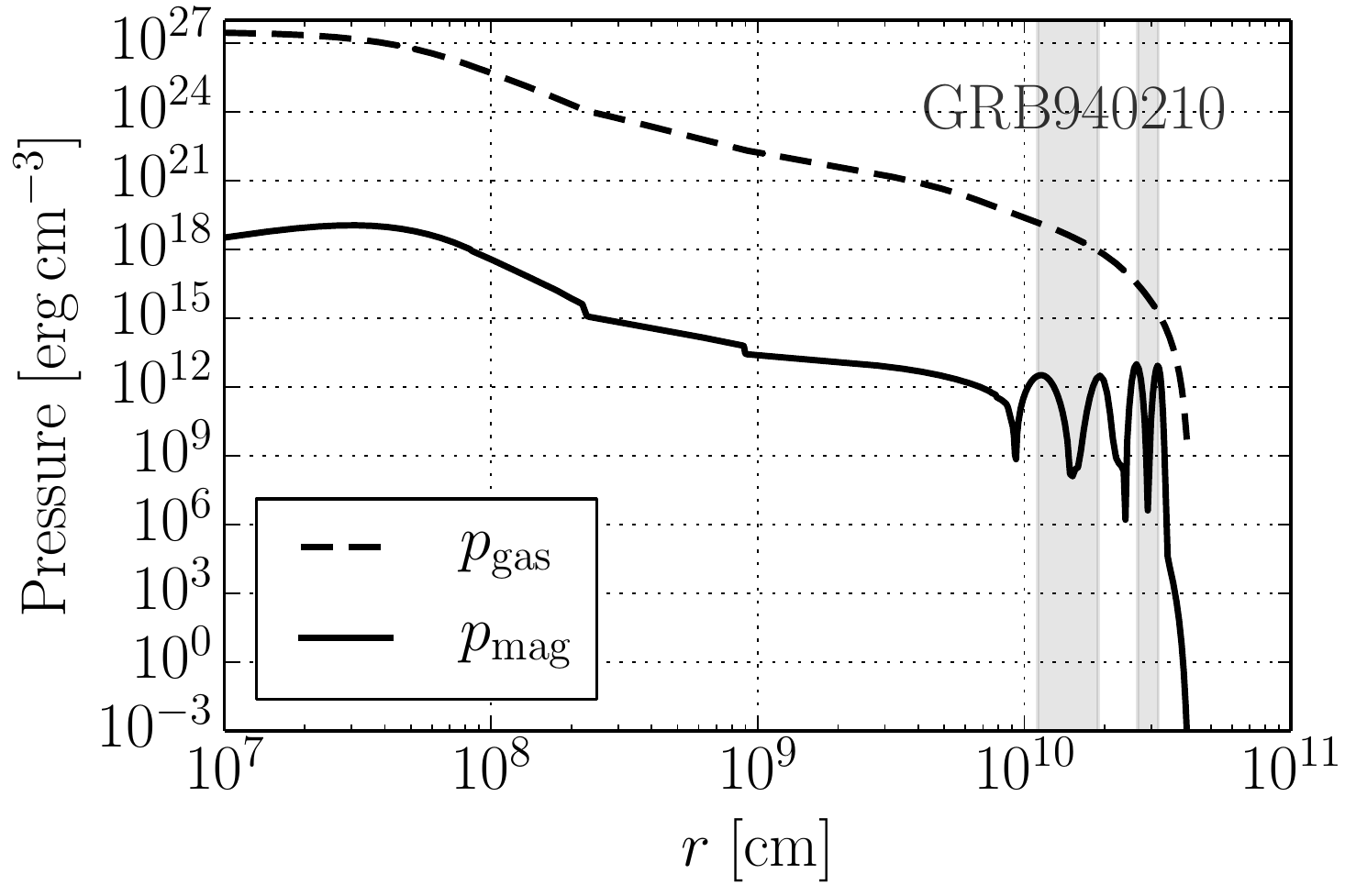}
\end{center}
  \caption{The radial profile of gas and magnetic pressures
    corresponding to Fig.~\ref{fig:lc2812knots2}. The
    magnetic pressure in the second knot is higher than that in
    Fig.~\ref{fig:lc2812knotsp}, reflecting a larger magnetic flux
    contained in that knot.}
\label{fig:lc2812knots2p}
\end{figure}

More generally, we suggest that rather long periods of silence between
bursts of prompt emission, such as in the case of GRB 940210 and GRB
920513, could be an imprint of large-scale magnetic field
inhomogeneities in the progenitor star. This opens up the possibility of stellar magnetic
flux ``tomography'': inferring the
magnetic flux distribution in progenitor stars by inverting the
information contained in their GRB light curves, as we have illustrated
for GRB 920513 and GRB 940210.
 
Note that, as seen in Fig.~\ref{fig:lc2812knots}, the GRB power does not strongly change from one interval of
activity to another (i.e., the cumulative light curve slope is
approximately the same among all active intervals).  But it 
changes from one quiescent interval to another (cumulative
light curves have different slopes in different quiescent
intervals). This behaviour is common to many GRBs
\citep{2002A&A...393L..29M}, and our model provides a natural
explanation for this: GRB power during quiescent intervals depends on
the magnetic flux in the knots, whereas the power during active
intervals is set by the net magnetic flux through the star and is
therefore roughly constant throughout the GRB. This is consistent with
suggestions \citep{2002MNRAS.331...40N} that quiescent intervals
follow a different type of time-variability than (i) short time-scale,
pulse-to-pulse variability of GRBs and (ii) long time-scale
variability set by the duration of the GRB.  Within our
model, the latter is set by the
free-fall time-scale of the outer layers of the star. The former could be caused by, e.g., the emission mechanism \citep{2009MNRAS.395..472K,2009MNRAS.394L.117N},
jet propagation effects \citep{2010ApJ...723..267M}, or the
variability of the central engine. We discuss these possibilities in
more detail at the end of this section.

In both cases of GRB 920513 and GRB 940210 the values of the $f-$factor
(see eqs.~\ref{eq:PhiBHtorus} and \ref{eq:PhiBHtorus2}), which sets
the depth of the dip in the light curve due to the accretion of a
magnetic knot, are of order of unity. This is precisely what is needed
for the quiescent interval to be noticeable: in order to substantially
modify the light curve, the knot must contain a magnetic flux
comparable or larger than that on the BH, $f\gtrsim1$. For this
reason, it is surprising that for both of the above GRBs we find
$f<1$, i.e., the flux in the knot is lower than the net flux through the
progenitor star. How does a knot know about the net magnetic flux
through the star?

We tried and were unable to obtain a visually good fit to the GRB
920513 lightcurve with $f>1$. The reason for this is clear. For a knot
with $f > 1$, the magnetic flux on the BH would vanish twice: once at
the beginning and once at the end of the consumption of the knot by
the BH.  This would lead to the cumulative GRB light curve that
flattens at the beginning and at the end of the quiescent interval and
steepens in between. No such behaviour is obviously present in
Fig.~\ref{fig:lc1606}: in fact, the cumulative light curve appears to
rise continuously, suggesting that the knot that led to the quiescent
interval in GRB 920513 light curve contained a relatively small
magnetic flux compared to that on the BH. The situation is different
with GRB 940210, whose light curve is shown in
Fig.~\ref{fig:lc2812knots}.  The light curve appears to show precisely
the type of behaviour indicative of a large flux in the knot, at least
for the second quiescent interval of the two. Indeed, we find that
$f_2 = 1.68>1$ gives a rather good fit to the cumulative light curve,
as seen in Fig.~\ref{fig:lc2812knots2}, with the rest of the
parameters similar to those used for the fit shown in
Fig.~\ref{fig:lc2812knots}: $f_1 = 0.85$, $t_1 = 6.2$~s, $\Delta t_1 =
3.1$~s, $t_2 = 18.7$~s, and $\Delta t_2 = 3.1$~s. The difference in
magnetic flux is reflected in Figure~\ref{fig:lc2812knots2p}, which
shows that the magnetic pressure in the second knot (indicated with
the right-most vertical grey stripe) is higher than in that in
Fig.~\ref{fig:lc2812knotsp}.  The model with $f_2 \simeq 1.5$, shown
in Fig.~\ref{fig:lc2812knots2}, appears to give a visually better fit
than that with $f_2 \simeq 0.5$, shown in Fig.~\ref{fig:lc2812knots}.
This suggests that stellar magnetic inhomogeneities contain
magnetic fluxes larger than the net flux through the
star.

In the above, we assumed that the polarity of magnetic flux in the
knots is negative, i.e., once a knot is accreted, it leads to a
depression in BH magnetic flux and a quiescent interval in the GRB
light curve. However, it is possible that in some cases the polarity
of magnetic flux in the knots is positive. This then would lead to
``hyper-active'' intervals and substantial differences in GRB
luminosity from one active interval to another.  Statistical analysis
of GRB light curves can potentially be used to determine the rates of
occurrence of the two polarities of the magnetic knots, thereby
providing us with valuable and otherwise not directly accessible information
on the magnetic field structure inside of GRB progenitor stars.

So far we have considered magnetic flux inhomogeneities over two
different scales inside the progenitor star: large-scales, comparable
to $r_*$ (net stellar flux), and smaller (but still large) scales,
around $0.1 r_*$ (magnetic knots). It is conceivable that spatial
distribution of stellar magnetic flux follows a continuous
distribution. This spatial distribution of magnetic flux in a
progenitor star leads to temporal distribution of quiescent
intervals. Thus, temporal analysis of GRB lightcurves is warranted to
constrain this distribution and thereby to improve our understanding of
the geometry of magnetic fields in the progenitors of GRBs.

GRB light curves show variability over a wide range of time scales,
from milliseconds to tens and hundreds of seconds.
Until now, we have focused on how the properties of a GRB 
evolve on a timescale similar to its overall duration ($t_{\rm GRB}\sim
30$~s). We showed that the rotation of the star and the free-fall
time of its outer layers set the GRB duration. We suggested that the
spatial distribution of large-scale inhomogeneities of the magnetic
field in the outer layers of the star leads to quiescent intervals in
GRB light curves lasting $\gtrsim {\rm few}$~seconds.
For completeness, here we speculate on 
what drives faster GRB variability, e.g., GRB pulses on
$t_{\rm p}\lesssim 1$~s timescales.
Such GRB variability may originate from the jet interactions 
with the stellar material. If the collapsing star is not
axisymmetric, it can imprint perturbations, e.g., bends, onto the
jet. Alternatively, if the jet is strongly magnetised, current-driven
instabilities may lead to non-axisymmetric structures as the jet
propagates through the star \citep{2013ApJ...764..148L,2014arXiv1402.4142B}.  Beaming
effects can result in large temporal changes of the observed radiation
even for modest changes of the Lorentz factor or the opening angle of
the jet. Since the characteristic timescale for sound or Alfv\'en
waves to cross the jet is $\sim1$~sec, short timescale variability may,
in part, be due to the jet propagation through the collapsar.

Another process that may modulate the jet power on a short timescale 
is the excursions of the accumulated magnetic flux from the
BH into the inner
accretion disk and vice versa. The accretion disk may switch between a
neutrino-cooled disk and a thick, advective disk multiple times
during the GRB.  Such transitions can
have a profound effect on the jet power and GRB luminosity. Neutrino cooling results in a
rather thin disk in the inner $R_{\rm thin}\sim 30r_g$
\citep{chen_neutrino_cooled_2007}. When the inner disk is thin, 
most of BH magnetic flux may diffuse outwards throughout $R_{\rm
  thin}$, leaving only a
small fraction of the magnetic flux on the BH and powering the jet. On
the other hand, a thick disk can push most of the available flux into
the BH, which can result in flares. Incidentally, the accretion
timescale for the neutrino-cooled disk is a fraction of a second and may
be or relevance for the typical duration of such gamma-ray pulses.

\section{Conclusions}
\label{sec:conclusions}

GRBs are characterised by the sudden onset of emission
that lasts for $T_{\rm GRB}\sim 1$~min and is followed by a sharp
 turn-off. The time-averaged properties of the prompt emission 
such as luminosity do not show any clear systematic 
trends during the GRB. Within the collapsar framework, the 
trigger of the burst can be naturally associated with the moment 
at which the jet breaks through the surface of the collapsing star.
The trigger takes place at time $t_{\rm trigger}\sim 10$~s after core
collapse \citep{mac99,2014arXiv1407.0123B,2014arXiv1402.4142B}. 
Since $T_{\rm GRB}\gtrsim t_{\rm trigger}$,
one expects that the accretion rate $\dot M$ at the BH evolves
(drops) appreciably over $T_{\rm GRB}$. As a result, in any model for which the jet 
power directly depends on the accretion rate $\dot M$, one would expect the
burst to become fainter throughout its duration until it is not
detectable anymore. Such a behaviour is not observed.  

Associating the gravitational energy release during the 
accretion onto the BH with the power of the 
GRB jets is hard on energetic grounds. 
Wolf Rayet stars have their masses
varying over a narrow range. The black hole
grows by several solar masses over the first minute 
past core collapse with the accretion power released 
$\gtrsim 10^{54}$erg.  
GRB jets, on the other hand, come in a broad range of energies 
$\sim 10^{49-52}$erg.
Therefore, another parameter, and not the accreted mass, sets 
the GRB energetics and dispersion in their properties. We argue that
this parameter is stellar magnetic flux.

To show this, we explored a number of models for collapsing Wolf-Rayet
stars and demonstrated that, by the time the jet breaks through the
collapsing star and the GRB becomes visible, the mass and spin of the
BH are close to their asymptotic values, i.e., $\MBH\sim 10M_{\odot}$,
$a\simeq 1$, respectively. For reasonable assumptions about magnetic
flux distribution in the star, the same holds true for the total
magnetic flux $\PhiBH$ at the BH.  Since the jet power in the
Blandford-Znajek model depends only on these 3 weakly changing parameters ($\MBH,
a, \PhiBH$), this explains the rough constancy of GRB luminosity during
the burst. 

But what causes the sharp turn-off of the GRBs? Since mass accretion rate
decreases asymptotically to zero, eventually, the accretion rate drops
to such a low level that the disk cannot confine the BH magnetic flux
anymore. Then, the flux diffuses out into the disk, and the disk
enters the \emph{magnetically-arrested disk} (MAD) state (see
Sec.~\ref{sec:steep-decline-phase}). In MADs, jet power is
proportional to mass accretion rate, and the jet power drops fast with
time as $L_j\propto \dot M\sim t^{-2...-20}$. Thus, it
is the transition to the MAD regime causes the abrupt turnoff of the
GRBs seen with \emph{Swift}.

What leads to the large burst-to-burst variation of GRB luminosity? In
our model, the jet power is set by the magnetic flux available in the
progenitor star. Typical BH magnetic flux of $\PhiBH\sim
10^{27.5}$~G~cm$^2$ is required to account for the observed luminosity
of bright bursts, which implies a progenitor with surface magnetic
field strength of $B\sim 10^4$G. Given the large observed variation in
surface stellar magnetic field strength, it is conceivable that
$\PhiBH$ varies by a factor of $\gtrsim30$ between different
progenitors, and since jet power scales as $\PhiBH$ squared, such a
variation translates into 3 orders of magnitude variation in jet
power, which is more than sufficient to account to the full observed
range of GRB luminosities.

In addition to power, the GRBs show a large variety in their
durations, from a few to a hundred seconds. What can cause such
diversity? In our model, the GRB starts at the jet break out time and ends
abruptly when the disk enters the MAD state. The jet breakout time is
controlled by the longest of disk formation and jet propagation
times. In most cases, it takes $\sim$ten seconds for the disk to form,
which sets the trigger time, after which the GRB is detectable. It
takes longer for the disk to form in progenitor stars with slow
rotation. In such cases, the formation of the disk (and jet) can be
substantially delayed, by tens of seconds, thereby shortening the GRB
duration. Other than that, the slow rotation
does not leave noticeable imprint on the GRB light curves, thus
core-collapse GRBs with durations much shorter than the free-fall time
of the outer layers of the progenitor star (which is around $100$~s),
might be suggestive of low angular momentum of their progenitors.

What can cause slow stellar rotation? One possibility is that magnetic
torques couple the rotation of different layers inside the star, slow
down the rotation of the core and  potentially
lead to near solid-body rotation of the progenitor, with the outer
layers of the star carrying most of the stellar angular momentum. We
showed that even such an extreme scenario can be capable of leading
a long-duration GRB, suggesting that GRBs can be more robust than
previously thought.  Thus, in order to get a
successful long-duration GRB, the progenitor star needs to contain
sufficient angular momentum for the accretion flow to encounter a
centrifugal barrier and spin up the central BH to a moderate spin. 
If high metallicity is conducive to the
production of strong stellar winds, which remove stellar
angular momentum, low stellar metallicity might be conducive to the
production of long-duration GRBs.
In the limit of extremely low stellar angular
momentum, infalling stellar material does not encounter a centrifugal
barrier, and an accretion disk does not form. In this scenario, the
jet forms just before the onset of MAD, and we suggest that such
extreme cases can be counterparts of failed or low-luminosity GRBs. 
These considerations may indicate that the flux through the progenitor
is at least as a decisive of a factor for a successful GRB as
the rotation of the collapsing star. 

The MAD formation time (=the time at which GRB ends) is very close
to the free-fall time of the outer layers of the star. Thus, another
factor that controls the GRB duration is the radius of the progenitor
star, with larger stars leading to longer GRBs. In this work, we
neglected the effects of accretion disk formation on the GRB duration, 
which we estimate are insignificant, unless stellar rotation makes up
a substantial fraction of Keplerian velocity.  If the stellar
rotation is near-Keplerian, an accretion disk of a large size and long
accretion time can form, thereby extending the central engine activity
time beyond the free-fall time of the outer layers of the star.

We used our model to probe the structure of the magnetic flux in the
star. By fitting the lightcurve of GRB 949817, we infer
that the total magnetic flux through the BH varies little throughout
the burst. The same conclusion holds for the majority of bright GRBs
that exhibit approximately linear cumulative count curves
\citep{2002A&A...393L..29M}. We argue that quiescent intervals seen in
many GRB light curves arise due to inhomogeneities of the magnetic
flux distribution inside the star. In at least one case, GRB 920513,
there is a clear deviation from linear cumulative curve during the
second half of that GRB. We infer the presence of a magnetised torus,
or a ``knot'', in the outer half of the star. In another case, GRB
940210, which shows two such deviations, we infer the presence of two
such knots, with indications that the outermost knot contains a magnetic
flux that exceeds the net flux through the star. 
 While the magnetic field structure in the stellar
interiors is an open question, knot-like magnetic flux configurations
are expected from MHD stability arguments
\citep{2004Natur.431..819B,2006A&A...453..687B}. Statistical analysis
of a larger number of GRB lightcurves might be a powerful probe of
stellar magnetic structure.

\section*{Acknowledgments}
We thank Eliot Quataert, Brian Metzger, and Omer Bromberg 
for valuable discussions.
AT was supported by NASA through Einstein Postdoctoral Fellowship
grant number PF3-140115 awarded by the Chandra X-ray Center, which is
operated by the Smithsonian Astrophysical Observatory for NASA under
contract NAS8-03060, and NASA via High-End Computing (HEC) Program
through the NASA Advanced Supercomputing (NAS) Division at Ames
Research Center that provided access to the Pleiades supercomputer, as
well as NSF through an XSEDE computational time allocation
TG-AST100040 on NICS Kraken, Nautilus, TACC Stampede, Maverick, and
Ranch.  DG acknowledges support from the Fermi 6 cycle grant number
61122. We used Enthought Canopy Python distribution to
generate figures for this work.

\appendix

\section{Timescales involved in core-collapse and accretion}
\label{sec:timesc-involv-core}
\begin{figure}
\begin{center}
    \includegraphics[width=0.9\columnwidth]{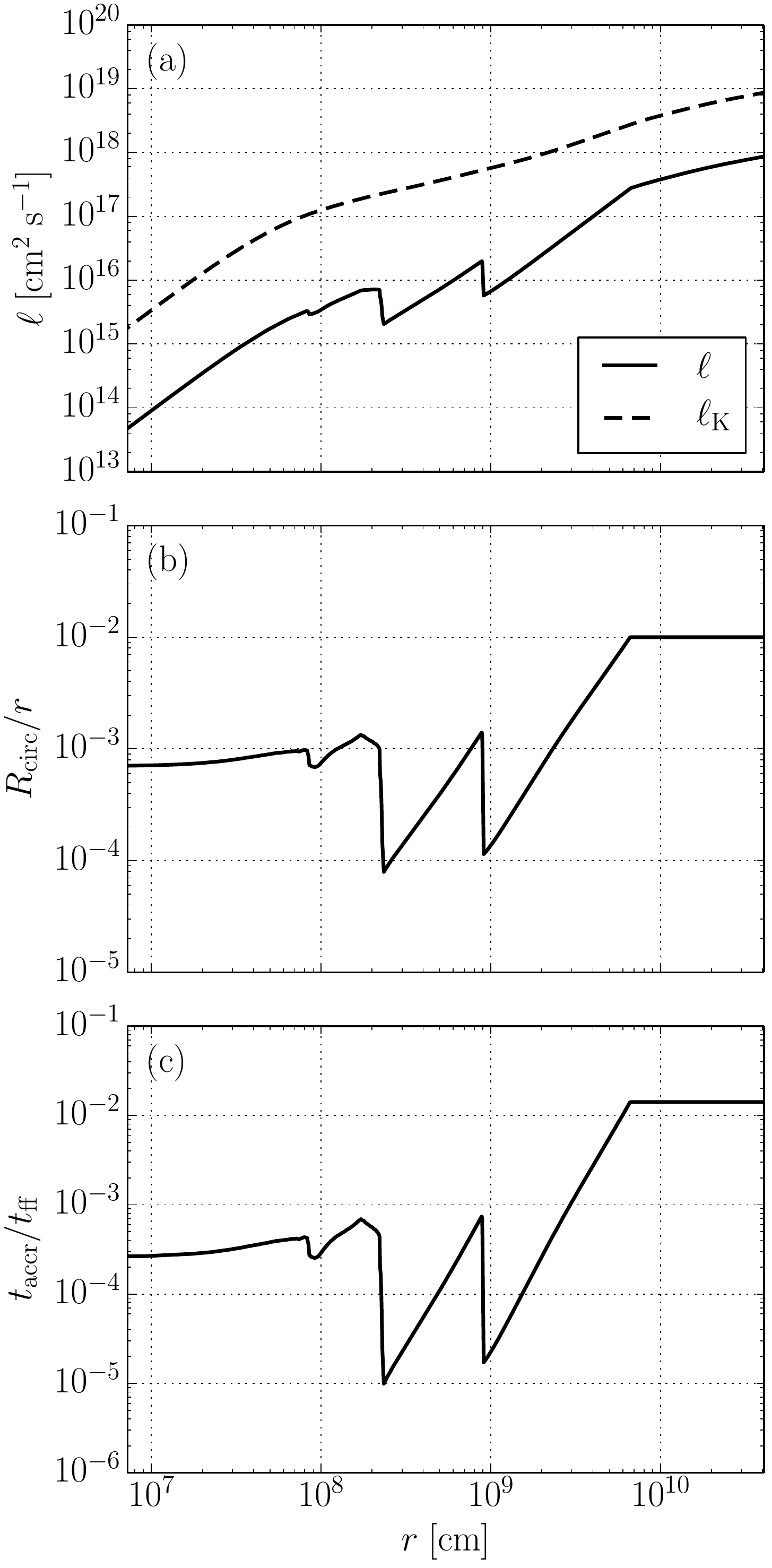}
\end{center}
\caption{The rotation and time-scales in model 16TI. As we discussed
  previously$^{\ref{fn:1}}$, rotation in progenitor models is uncertain,
  and in fact it can exceed the Keplerian value. Thus, in our work, we
  limit the rotation to be at most $10$\% of Keplerian value. The
  resulting angular momentum profile is shown in {\bf Panel (a)} with
  the solid line; the Keplerian angular momentum is shown with the
  dashed line. We assume that gas falls freely from
  its initial position $r$, until it
  hits the centrifugal barrier at the circularisation radius, $R_{\rm
    circ}$; the ratio of the two is shown in {\bf Panel (b)}. 
  Inside of that radius,
  gas travels viscously in the form of an accretion disk, with the
  accretion time $t_{\rm acc} \approx r/v_r$, that is
  much shorter than the free-fall time, $t_{\rm ff}$, as seen in
  {\bf Panel (c)}. In the plot, we set
  $h/r = 1$ and $\alpha = 0.1$. Since gas transit time though the
  accretion disk is negligible compared to the free-fall
  time, we for simplicity ignore it.}
\label{fig:timescales}
\end{figure}

As we discussed
previously$^{\ref{fn:1}}$, rotation in progenitor models is uncertain,
and in fact it can exceed the Keplerian value (see
Fig.~\ref{fig:rhoell}b). Thus, in our work, we
limit the rotation to be at most $10$\% of Keplerian value. The
resulting angular momentum profile is shown in
Fig.~\ref{fig:timescales}(a). 

We assume that gas falls freely from its initial position
$r$, until it hits the centrifugal barrier at the circularisation
radius, $r_{\rm D}$, 
\begin{equation}
\frac{r_{\rm D}}{r}=\left(\frac{\ell}{\ell_{\rm K}}\right)^2. \label{eq:Rcirc}
\end{equation}
The ratio \eqref{eq:Rcirc} is
shown in Fig.~\ref{fig:timescales}(b).  Inside of $r_{\rm D}$, gas travels
viscously in the form of an accretion disk, with the accretion time
\begin{equation}
t_{\rm acc} \approx \frac{r_{\rm D}}{v_r} \approx \alpha^{-1} \left(\frac{R_{\rm
  circ}}{r_g}\right)^{1/2}\left(\frac{h}{r}\right)^{-2}\frac{r_g}{c}.
\label{eq:tacc}
\end{equation}
Figure~\ref{fig:timescales}(c)
shows the ratio if accretion time to free-fall time, $t_{\rm ff}=(2r_g/r)^{1/2}$:
\begin{equation}
  \label{eq:taccotff}
  \frac{t_{\rm acc}}{t_{\rm ff}} = 
  \frac{2^{1/2}}{\alpha}
  \left(\frac{r_{\rm D}}{r}\right)^{3/2}
  \left(\frac{h}{r}\right)^{-2}
  =
  \frac{2^{1/2}}{\alpha}
  \left(\frac{\ell}{\ell_{\rm K}}\right)^{3/2}
  \left(\frac{h}{r}\right)^{-2}.
\end{equation}
In Fig.~\ref{fig:timescales}, for illustration we choose
$h/r = 1$ and $\alpha = 0.1$. Since gas transit time though the
accretion disk is negligible compared to the free-fall time, we for
simplicity ignore the accretion time. That said, there could be scenarios in which $t_{\rm acc} \gg
t_{\rm ff}$ (e.g., if stellar rotation is close to Keplerian).  In
such extreme cases, one does need to account for the accretion
time.

\bibliographystyle{mn2e} 

{\small

}

\label{lastpage}
\end{document}